\documentclass[12pt]{article}
\usepackage{graphicx}
\usepackage{enumerate}
\usepackage{natbib}
\usepackage{url} 

\usepackage{amsmath, amssymb}
\usepackage[colorlinks,citecolor=blue,urlcolor=blue]{hyperref}
\usepackage{enumitem}

\usepackage{multirow}
\usepackage{shortcuts}
\usepackage{graphicx}
\usepackage{nicefrac}
\usepackage{subcaption}
\usepackage{booktabs}
\usepackage{todonotes}
\usepackage[flushleft]{threeparttable}
\usepackage{bm}
\usepackage{bbm}
\usepackage[flushleft]{threeparttable}
\usepackage{float}
\usepackage{extarrows}

\let\given\givenbase

\usepackage{amsthm}

\usepackage{accents}
\newlength{\dhatheight}

\numberwithin{equation}{section}
\theoremstyle{plain}
\newtheorem{theorem}{Theorem}[section]
\newtheorem{assumption}{Assumption}[section]

\newtheorem{lemma}[theorem]{Lemma}
\theoremstyle{definition}

\usepackage[ruled]{algorithm2e}

\SetArgSty{textrm}  
\usepackage{algorithmic}

\newcommand{\Study}{{\mathcal S}}
\newcommand{\TT}{{\mathcal T}}
\newcommand{\TC}{{\mathcal C}}
\newcommand{\Unt}{{\mathcal U}}

\newcommand{\Xs}{{X_{1:n}}}
\newcommand{\Ts}{{T_{1:n}}}

\newcommand{\Ws}{{W_{1:n}}}
\newcommand{\Vs}{{V_{1:n}}}
\newcommand{\Ss}{{S_{1:n}}}
\newcommand{\lambdas}{{\lambda_{0:1}}}

\newcommand{\ie}{\emph{i.e.}}

\usepackage{xcolor}

\usepackage{multirow}

\newcommand{\blind}{1}

\begin{document}

\def\spacingset#1{\renewcommand{\baselinestretch}%
{#1}\small\normalsize} \spacingset{1}



\if1\blind
{
  \title{\bf Optimal Estimation of Generalized Average Treatment Effects using Kernel Optimal Matching}
  \author{Nathan Kallus \\
    Operations Research and Information Engineering \\ and Cornell Tech, Cornell University, New York, NY, USA\\
    and \\
    Michele Santacatterina\thanks{
    \textit{∗Corresponding author. This material is based upon work supported by the National Science Foundation under Grants Nos.  1656996 and 1740822. The Authors would like to thank Brenton Pennicooke for the access to the QOD dataset and Mattias Larsson, Anders S{\"o}nnerborg, Chuc Nguyen Thi Kim, Do Duy Cuong and Tam V Vu for the access to the HIV trial data on peer support. The authors also thanks the Multicenter AIDS Cohort Study consortia for the access to their public data set. }}\\ 
    Tripods Center for Data Science for Improved Decision Making \\ and Cornell Tech,  Cornell University,  New York, NY, USA}
  \maketitle
} \fi

\if0\blind
{
  \bigskip
  \bigskip
  \bigskip
  \begin{center}
    {\LARGE\bf 
Optimal Estimation of Generalized Average Treatment Effects using Kernel Optimal Matching}
\end{center}
  \medskip
} \fi

\bigskip
\begin{abstract}
In causal inference, a variety of causal effect estimands have been studied, including the sample, uncensored, target, conditional, optimal subpopulation, and optimal weighted average treatment effects. Ad-hoc methods have been developed for each estimand based on inverse probability weighting (IPW) and on outcome regression modeling, but these may be sensitive to model misspecification, practical violations of positivity, or both.
The contribution of this paper is twofold. First, we formulate the generalized average treatment effect (GATE) to unify these causal estimands as well as their IPW estimates. Second, we develop a method based on Kernel Optimal Matching (KOM) to optimally estimate GATE and to find the GATE most easily estimable by KOM, which we term the Kernel Optimal Weighted Average Treatment Effect.
KOM provides uniform control on the conditional mean squared error of a weighted estimator over a class of models while simultaneously controlling for precision. We study its theoretical properties and evaluate its comparative performance in a simulation study. We illustrate the use of KOM for GATE estimation in two case studies: comparing spine surgical interventions and studying the effect of peer support on people living with HIV.
\end{abstract}

\noindent%
\vfill

\newpage
\spacingset{1.5} 

%
%
\section{Introduction}
        \label{sec:intro}
        
        One of the primary goals of causal inference is to estimate the average causal effect of a treatment or intervention on an outcome under study. A common causal estimand of interest is the Sample Average Treatment Effect (SATE), which is the average effect of a treatment on an outcome among all individuals in the sample. 
        Often, however, we may be interested in other averages. For example, 
        \citet{stuart2010,buchanan2018generalizing} consider the Target Average Treatment Effect (TATE) on a population or sample distinct from the study sample and propose the use of inverse probability of sampling weights. 
        Similarly, if outcome data are only available for some units, \citet{cain2009inverse,robins2000correcting} propose the use of inverse probability of censoring weights to generalize the results to the whole sample.
        Other estimands of interest focus on particular subgroups of the sample such as the Sample Average Treatment Effect on the Treated (SATT), the Conditional Average Treatment Effect (CATE) \citep{crump2008nonparametric,cai2010analysis}, and the Complete-Case SATE (CCSATE) \citep{seaman2013review}.
        In particular, \citet{crump2009dealing} propose the Optimal SATE (OSATE) and, as in \cite{li2018balancing}, the Optimal Weighted Average Treatment Effect (OWATE) as the average treatment effect restricted by or weighted by overlap in covariate distributions in order to make the estimation easier. 
        
        Ad-hoc methods, such as those bases on Inverse Probability Weighting (IPW) \citep{horvitz1952generalization,robins1994estimation,robins2000marginal,lunceford2004stratification}  and outcome regression modeling, have been widely used to estimate these causal estimands. However, due to their sensitivity to model misspecification these methods may lead to biased estimates. 
        In addition, IPW-based methods depend heavily on the positivity assumption, which practical violations of lead to extreme weights and high variance \citep{robins1995analysis,scharfstein1999adjusting,robins2007comment,kang2007demystifying}. In Section \ref{related_work} in the Supplementary Material, we thoroughly discuss these issues, some of the related work to overcome them and alternative methodologies to estimate the aforementioned causal estimands. 
        
        In this paper, we start by presenting a general causal estimand, the Generalized Average Treatment Effect (GATE), which unifies all the causal estimands previously presented and motivates the formulation of new ones.  We then present and apply Kernel Optimal Matching (KOM) \citep{kallus2016generalized,kallus2018more} to optimally estimate GATE. KOM provides weights that simultaneously mitigates the possible effect of model misspecification and control for possible practical positivity violations \citep{kallus2018more}. We do that by minimizing the worst-case Conditional Mean Squared Error (CMSE) of the weighted estimator in estimating GATE over the space of weights. The proposed methodology has several attractive characteristics. First, KOM can be used to optimally estimate a variety of well-known causal estimands, as well as to find new ones such as the Kernel Optimal Weighted Average Treatment Effect (KOWATE). In Section \ref{mincmse} we show that various causal estimands can be easily estimated by simply modifying the optimization problem formulation we give for KOM, which is fed to an off-the-shelf solver. Second, by minimizing the worst-case CMSE of the weighted estimator, it leads to better accuracy, precision, and total error. We show this in our simulation study in Section \ref{simu}. Third, by optimally balancing covariates, KOM mitigates the effect of possible model misspecification. In Section \ref{simu}, we show that both absolute bias and root MSE (RMSE) of the weighted estimator that uses weights obtained by using KOM are consistently lower across levels of misspecification. Fourth, by penalizing the weights, KOM controls precision. We show this in Section \ref{simu}. Fifth, the weights are obtained by using off-the-shelf solvers for convex-quadratic optimization. Finally, KOM is implemented in an open source \texttt{R} package.
        
        In the next Section we introduce notation, specify assumptions and define GATE, the estimand of interest and its weighted estimator. We then introduce KOM for GATE, describe its theoretical properties and present some practical guidelines on its use (Section \ref{kom}). In Section \ref{simu}, we present the results of a simulation study aimed at comparing the performance of KOM with IPW, overlap weights, truncated weights and outcome regression modeling with respect to absolute bias and RMSE across levels of practical positivity violations and levels of misspecification. In Section \ref{illustrations}, we apply KOM on the evaluation of the effect of spine surgical interventions on the Oswestry Disability Index (ODI) among patients with lumbar stenosis or lumbar spondylolisthesis, and on the evaluation of peer-support on CD4 cell count in two target populations of healthier patients, using real-world data. We conclude with some remarks in Section \ref{conclusions}.

%
%

\section{Generalized Average Treatment Effect}
        \label{gate}
        
        Suppose we have a simple random sample with replacement of size $n$ from a population. Under the potential outcome framework \citep{imbens2015causal}, for each unit $i=1,\dots,n$, we let $Y_i(t) \in \Rl$ be the potential outcome of treatment $t \in \lbrace 0,1\rbrace$. We let $X_i \in \mathcal X$ be the observed confounders. We consider three exclusive and exhaustive subsets of the units:  (i) $i\in\TT$ units treated with $t=1$, for whom we observe $Y_i=Y_i(1)$; (ii) $i\in\TC$ units treated with $t=0$, for whom we observe $Y_i=Y_i(0)$ and (iii) $i\in\Unt$ untreated units, for whom we do not observe anything but confounders.  We let $\Study=\TT\cup\TC$, be the units in the study sample.  We set $T_i=\indic{i\in\TT}$, the indicator of being treated with $t=1$, $S_i=\indic{i\in\Study}$, the indicator of being in the study sample, and $U_i=\indic{i\in\Unt}=1-S_i$, the indicator of being outside the study. 
        Let $g_t(X)=\Eb{Y_i(t)\mid X_i, S_i=1}$, for $t \in {0,1}$. We define the Generalized Average Treatment Effect (GATE)  as the weighted average difference between the conditional expectation of the potential outcome of those treated and those untreated conditioned on $\Xs$. Formally, we define GATE as,
        \begin{equation}
        \label{GATE_estimand}
            \tau_V=\frac{1}{n} \sum_{i=1}^n V_i (g_1(X_i)-g_0(X_i)),
        \end{equation}
        
        \noindent
        where $V_i$ is chosen to target the estimand of interest and may depend on $X_{1:n},T_{1:n},S_{1:n}$ (see Assumption~\ref{honest} below). For instance, when $V_i=S_i$, we target the study SATE, and when $V_i=\frac{nU_i}{\abs{\Unt}}$, we target the TATE
        .  Moreover, by setting $V_i$ equal to the overlap weights \citep{li2018balancing} and the truncated weights \citep{crump2009dealing}, we target the OWATE and OSATE, respectively. We provide examples of causal estimands in the first two columns of Table~\ref{table1}. To estimate GATE in eq.~\eqref{GATE_estimand} we propose to use the following weighted estimator,
        \begin{equation}
        \label{GATE_westimator}
            \hat\tau_W=\frac{1}{n}\sum_{i\in\TT}W_iY_i-\sum_{i\in\TC}W_iY_i=\frac{1}{n}\sum_{i\in\Study}W_i(-1)^{(T_i+1)}Y_i.
        \end{equation}
        
        For instance, the usual IPW estimator for SATE is given by plugging in $W_i=W_i^{\op{IPW}}=\frac{T_i}{\phi(X_i)}+\frac{1-T_i}{1-\phi(X_i)}$, where $\phi(X_i)=\Prb{T_i=1\mid X_i=x,S_i=1}$, is the propensity score. In the next Lemma,  we provide a general formulation of the IPW weights that make $\hat{\tau}_{W^{\op{IPW}}}$ unbiased for GATE for any $\Vs$.
        To do so, we impose the assumption of consistency, non-interference, ignorable treatment assignment and ignorable sample assignment \citep{imbens2015causal}. Consistency, states that the observed outcome corresponds to the potential outcome of the treatment applied to that unit, and non-interference reflects the fact that units potential outcomes are not effected by how the treatment or intervention has been allocated. Consistency together with non-interference are also known as SUTVA  \citep{imbens2015causal}. 
        Ignorable treatment assignment, (also called unconfoundeness, no unmeasured confounding, or exchangeability), states that the potential outcome, $Y_i(t)$, is independent to the treatment assignment mechanism given covariates. Similarly, ignorable sample assignment states that the potential outcome, $Y_i(t)$, is independent to the sampling assignment mechanism, \ie,~ being part of the study sample, given covariates. We formalize these assumptions as follow,  
        
        \begin{assumption}[Ignorable treatment assignment]\label{unconftrt}
        $Y_i(t)\indep T_i\mid X_i,S_i=1$
        \end{assumption}
        
        \begin{assumption}[Ignorable sampling]\label{unconfinc}
        $Y_i(t)\indep S_i \mid X_i$
        \end{assumption}
        
        \begin{assumption}[Boundedness of $\phi(X_i)$]\label{bound1} The propensity score $\phi(X_i)=\Prb{T_i=1\mid X_i,S_i=1}$ is bounded away from 0,1.
        \end{assumption}
        \begin{assumption}[Boundedness of $\psi(X_i)$]\label{bound2} The sampling probability  $\psi(X_i)=\Prb{S_i=1\mid X_i}$ is bounded away from 0.
        \end{assumption} 
        
        Letting $H_{1:n} = \lbrace X_{1:n},T_{1:n},S_{1:n} \rbrace$,
        we additionally assume, 
        
        \begin{assumption}[Honest weights]\label{honest}{ $\Ws$ and $\Vs$ are independent of all else given $H_{1:n}$.}
        \end{assumption}

        In the next Lemma we define the genalized IPW weights, $W_{1:n}^{\op{IPW}}$, and show  that $\hat{\tau}_{W^{\op{IPW}}}$, the weighted estimator in eq.~\eqref{GATE_westimator} weighted by $W_{1:n}^{\op{IPW}}$, is unbiased for GATE.
        
        \begin{lemma}
        \label{lemma1}
         Define 
        \begin{align*}
            W_i^{\op{IPW}}= \frac{1}{\psi(X_i)} \prns{\frac{T_i}{\phi(X_i)} + \frac{1-T_i}{(1-\phi(X_i))} }& \left(  \phi(X_i) V_i(\Xs,(T_{-i},1),\Ss) \right. \\ 
            &+ \left. (1-\phi(X_i)) V_i(\Xs,(T_{-i},0),\Ss)  \right),
        \end{align*}
        \noindent
        where $(T_{-i},t)$ is equal to $\Ts$ in all components except the $i$-th where it is equal to $t$. 
        Then under consistency, non-interference and assumptions \ref{unconftrt}-\ref{honest}, $$\mathbbm{E}\left[{\hat{\tau}_{W^{\op{IPW}}} - \tau_V} \given  H_{1:n} \right] =0.$$
        \end{lemma}
        
        This is a well-known results for SATE \citep{lunceford2004stratification} and TATE \citep{stuart2011use,buchanan2018generalizing}. 
        If we assume appropriate bounds on the norms of $V$ and the variances of $Y_{1:n}$, it is easy to additionally see that $\hat{\tau}_{W^{\op{IPW}}}$ has diminishing variance and is therefore also consistent.
        We show examples of inverse probability weights, $W_i^{\op{IPW}}$, for the IPW estimator of GATE in the last column of Table~\ref{table1}. 
        
        \begin{table}[t!]
        \centering
          \begin{threeparttable}
        \caption{Examples of causal estimands, the corresponding weights $V_i$ of eq.~\eqref{GATE_estimand}, and inverse probability weights, $W_i^{\op{IPW}}$ for the IPW estimator. \label{table1}}
        \begin{tabular}{ccc}
        \hline
        Estimand & $V_i$                            & $W_i^{\op{IPW}}$                                                    \\ 
        \hline
        SATE     & $S_i$ & $\frac{T_i}{\phi(X_i)} + \frac{1-T_i}{(1-\phi(X_i))} $                          \\
        SATT     & $\frac{n T_i}{\sum_j^n T_j}$ & $\frac{T_i}{\sum_j^n T_j} + \frac{(1-T_i)}{(1-\phi(X_i))} $                          \\
        TATE     & $\frac{nU_i}{\abs{\Unt}}$ & $\frac{S_i}{\abs{\Unt} \psi(X_i)}\prns{\frac{T_i}{\phi(X_i)} + \frac{(1-T_i)}{(1-\phi(X_i))} }$                    \\
        OWATE   & $n \frac{\phi(X_i)(1-\phi(X_i))}{n_{\op{O}}}$ &  $\frac{1}{n_{\op{O}}}\prns{
        T_i + (1 - 2 T_i) \phi(X_i)
        }$                                                        \\
        OSATE    & $n \frac{I_{\alpha}[\phi(X_i)]}{n_{\op{T}}}$          &      $\frac{I_{\alpha}[\phi(X_i)]}{n_{\op{T}}} \prns{ \frac{T_i}{\phi(X_i)} + \frac{1-T_i}{(1-\phi(X_i))} }$ \\
        \hline
        \end{tabular}
        \begin{tablenotes}
              \small
              \item Notes:  $\phi(X_i)=\Prb{T_i=1\mid S_i=1,X_i}$, is the propensity score,  $\psi(X_i)=\Prb{S_i=1\mid X_i}$ is the probability of being in the sample, $n_{\op{O}}=\sum_{i=1}^n\phi(X_i)(1-\phi(X_i))$, $n_{\op{T}}=\sum_{i=1}^nI_{\alpha}[\phi(X_i)]$, and $I_{\alpha}[\phi(X_i)]=\indic{\alpha<\phi(X_i)<1-\alpha}$.
            \end{tablenotes}
          \end{threeparttable}
        \end{table}

        In the next Section, we introduce Kernel Optimal Matching for estimating GATE, which, instead of plugging estimated propensities into the weighted estimator,  provides weights that minimizes the CMSE of $\hat\tau_W$ for GATE. By doing so, the proposed methodology optimally minimizes the bias with respect to GATE while simultaneously controlling precision. We further consider simultaneously choosing $V$ to minimize the worst-case CMSE to obtain KOWATE.

%
%

\section{Kernel Optimal Matching for estimating GATE}
        \label{kom}
        
        In this Section, we present Kernel Optimal Matching for estimating GATE. We start by decomposing the CMSE of the weighted estimator, $\hat\tau_W$, in eq.~\eqref{GATE_westimator}. We show that this CMSE can be decomposed in terms of (a) the discrepancies between the conditional expectation of the potential outcome among the treated and the control, and (b) a variance term (Section \ref{deco}).  Since the CMSE depends on some unknown functions (conditional expectations), in Section \ref{wccmse}, we guard against all possible realizations of the unknown functions by considering the worst-case CMSE of $\hat\tau_W$.  In Section \ref{mincmse}, we embed these in reproducing kernel Hilbert spaces (RKHS) and use quadratic programming to minimize the corresponding worst-case CMSE and find optimal weights.

\subsection[deco]{Decomposing the CMSE of $\hat\tau_W$}
        \label{deco}
        
        We now decompose the CMSE of $\hat\tau_W$, the weighted estimator for GATE. Recall that, in Section \ref{gate}, we defined $g_t(X)=\Eb{Y_i(t)\mid X_i, S_i = 1}$. Further define $\epsilon_{it}=Y_i(t)-g_t(X_i)$ and $\sigma^2_{it}=\op{Var}\prns{Y_i(t)\mid X_i, S_i = 1}=\Eb{\epsilon_{ti}^2\mid X_i, S_i = 1}$, for $t \in {0,1}$, and $\sigma_i^2=T_i\sigma^2_{i1}+(1-T_i)\sigma^2_{i0}$. We then define, for each function $f$, the $f$-moment discrepancy between the weighted $t$-treated study sample and the $V$-weighted total sample,
        $$
            B_t(\Ws,\Vs,f) = \frac{1}{n}\sum_{i=1}^n \left( S_i \indic{T_i=t} W_i - V_i \right) f(X_i),
        $$
        \noindent
        where $\indic{T_i=t}$ is equal to 1 if $T_i=t$ and 0 otherwise.  In the following theorem, we show that the CMSE of $\hat\tau_W$ can be decomposed into the squared such discrepancies in the conditional expectations of the potential outcomes.
        
        \begin{theorem}
        \label{thm1} Under consistency, non-interference and assumptions \ref{unconftrt}-\ref{honest}, 
        \begin{align}
        	\mathbbm{E}\left[\hat{\tau}_{\text{W}} - \tau_V \given  H_{1:n} \right] &=  B_1(\Ws,\Vs,g_1) - B_0(\Ws,\Vs,g_0)
        	\notag\\
        	\mathbbm{E}\left[\prns{\hat{\tau}_{\text{W}} - \tau_V}^2 \given  H_{1:n} \right] &= 
        	\left(B_1(\Ws,\Vs,g_1) - B_0(\Ws,\Vs,g_0)\right)^2 + \frac{1}{n^2}\sum_{i = 1}^n S_iW_i^2 \sigma_i^2. \label{thmcmse_cgate}
        \end{align}
        \end{theorem}

        In the next Section, we show how to find weights that minimize eq.~\eqref{thmcmse_cgate}. The main challenge in this task is that the functions $g_t$, on which this quantity depends, are unknown.

%
%

\subsection{Worst-case CMSE}
        \label{wccmse}
        
        To overcome the issue that we do not know the $g_t$-functions which the CMSE of $\hat\tau_W$ depends, we will guard against any possible realizations of the unknown functions. Specifically, since the CMSE of $\hat \tau_W$ scales linearly with $g_0$ and $g_1$, we consider its magnitude with respect to that of $g_0$ and $g_1$. We therefore need to define a magnitude. We choose the following,
        $$
        \fmagd{g} = \sqrt{ \fmagd{g_0}_0^2 + \fmagd{g_1}_1^2 },
        $$
        where $\fmagd{\cdot}_t$ are some extended seminorm on functions from the space of confounders to the space of outcomes. We discuss a specific choice of such extended seminorms in Section \ref{mincmse}. Given this magnitude, we can define the \textit{worst-case squared bias} as follows:
        \begin{align}
            \label{nbias}
        \mathcal{B}(\Ws,\Vs)&=  \sup_{g}\frac{B_1(\Ws,\Vs,g_1)-B_0(\Ws,\Vs,g_0)}{\|g\|} \\
        &= \sqrt{\Delta^2_1(\Ws,\Vs) + \Delta^2_0(\Ws,\Vs)},
        \end{align}
        where 
        $$
        \Delta_t(\Ws,\Vs)=\sup_{g_t} \frac{B_t(\Ws,\Vs,g_t)}{\|g_t\|_{t}}=\sup_{\|g_t\|_{t} \leq 1} B_t(\Ws,\Vs,g_t),
        $$
        is the \textit{worst-case discrepancy} in the $g_t$-moment between the weighted $t$-treated group and the $V$-weighted sample over all $g_t$ functions in the unit ball of $\|\cdot\|_t$. In particular, given a positive semidefinite (PSD) kernel $\mathcal K_t(x,x')$, if we choose the corresponding RKHS (a Hilbert space of functions with continuous evaluations, which is associated with the reproducing kernel $\mathcal{K}_t$) to specify the norm, we can show that the worst-case discrepancy can be expressed as a convex-quadratic function in $\Ws$. 
        
        \begin{theorem}
        \label{thm2}
        Define the matrix $K_t\in\mathbb R^{n\times n}$ as $K_{tij}=\mathcal K_t(X_{i},X_{j})$ and note that it is positive semidefinite by definition. Then,
        \begin{equation*}
        \begin{aligned}
            \Delta_t(\Ws,\Vs)^2 &= \frac{1}{n^2}\prns{\Ws^TI_{S}I_{t}K_tI_{S}I_{t}\Ws-2\Vs^TK_tI_{S}I_{t}W_{1:n}+\Vs^TK_t\Vs},
        \end{aligned}
        \end{equation*}
        \noindent
        where $I_{t}$ is the diagonal matrix with $\mathbb I[T_{i}=t]$ in its $i^\text{th}$ diagonal entry, and $I_{S}$ is the diagonal matrix with $\mathbb I[S_{i}=1]$ in its $i^\text{th}$ diagonal entry.
        \end{theorem}

        Based on Theorem \ref{thm2}, letting the RKHS given by the kernel $\mathcal{K}_t$ specify the norm, both the worst-case bias and the worst-case CMSE of $\hat \tau_W$ are convex-quadratic functions in $\Ws$. Specifically, we define the worst-case CMSE as
        \begin{align}
        \notag
        \mathfrak{C}(\Ws,\Vs,\lambdas) &= \sup_{\|g\|^2\leq1}
        \mathbbm{E}\left[\prns{\hat{\tau}_{\text{W}} - \tau_V}^2\mid H_{1:n}\right]\\\label{worstCMSE}&=\Delta^2_1(\Ws,\Vs)+\Delta^2_0(\Ws,\Vs)+\frac{\lambda_0}{n^2} \| I_SI_0 \Ws \|_2^2+\frac{\lambda_1}{n^2} \| I_SI_1 \Ws \|_2^2 ,
        \end{align}
        where, for simplicity, we use within-treatment-group equal variance weights, $\lambda_0,\lambda_1$. More generally, we can use any positive definite matrix $\Lambda$ to penalize the variances as $\Ws^T I_S\Lambda I_S \Ws$. In the next Section we show how to minimize the worst-case CMSE  of $\hat \tau_W$ in estimating GATE, $\mathfrak{C}(\Ws,\Vs,\lambdas)$, by using off-the-shelf solvers for quadratic optimization.

%
%

\subsection{Minimizing the worst-case CMSE}
        \label{mincmse}
        
        In the previous two Sections, we showed that the CMSE of $\hat\tau_W$ in estimating GATE can be decomposed in squared bias plus its variance. We also showed that, since the bias depends on unknown conditional expectations, by guarding against any possible realizations of these unknown functions, embedded in an RKHS given by the kernel $\mathcal{K}_t$, the worst-case CMSE of  $\hat\tau_W$ can be expressed as a convex-quadratic function in $\Ws$. Here, we use quadratic programming to obtain the weights $\Ws$ that minimizes the worst-case CMSE of $\hat\tau_W$. When interested in estimating, for example,  SATE, and TATE, the set of weights $\Vs$ is fixed, \ie, all $V_i$ are given, known scalars. We show the corresponding convex-quadratic optimization problem when the set of weights $\Vs$ is fixed in the next Section. In addition, given the flexibility of the proposed methodology, we can also let $\Vs$ be variable and let it be chosen by the solver in such a way that the worst-case CMSE of $\hat\tau_W$ is minimized. We show this in Section \ref{variableV}. 
        
        \subsubsection{Fixed $\Vs$}
        \label{fixedV}
        Let $\mathcal W=\fbraces{W_{1:n}\in\R n:W_i\geq0\;\forall i,\,\sum_{i\in\mathcal T}W_i=\sum_{i\in\mathcal C}W_i=n}$.  
        When $\Vs$ is fixed, we propose to use weights $\Ws$ obtained by solving the following optimization problem
        \begin{align}
        \underset{\Ws \in \mathcal{W}}{\min}
        \prns{\Delta^2_1(\Ws,\Vs)+\Delta^2_0(\Ws,\Vs)+\frac{\lambda_0}{n^2} \| I_SI_0 \Ws \|_2^2+\frac{\lambda_1}{n^2} \| I_SI_1 \Ws \|_2^2 }
        ,\label{kom_cmse1}
        \end{align}
        
        \noindent
        where $\lambda$ is interpreted as a penalization parameter that controls the trade-off between bias and variance. When $\lambda$ equals zero, we obtain weights that yield minimal bias. When $\lambda \rightarrow \infty$, we obtain uniform weights. (If we have estimates of heteroskedastic conditional variance, we can also easily use unit-specific weights.) We discuss how to tune this hyperparameter in Section \ref{guidelines}.
        As shown in Theorem \ref{thm2}, using an RKHS norm, we can show that the optimization problem \eqref{kom_cmse1} reduces to the following linearly-constrained convex-quadratic optimization problem:
        
        \begin{equation}
        \label{qpkom1}
        \begin{aligned}
        \underset{\substack{W_{1:n}\geq0,\\W_{1:n}^TI_{S}I_1e_n=n,\\W_{1:n}^TI_{S}I_0e_n=n}}{\min} \frac{1}{n^2} \prns{
        \Ws^T \prns{\sum_{t \in \{0,1\}}I_{S}I_{t} (K_t+\lambda_tI) I_{t}I_{S}}\Ws 
        -2\Vs^T( K_1I_{S}I_{1}+K_0I_{S}I_{0})\Ws}.
        \end{aligned}
        \end{equation}
        
        \subsubsection{Variable $\Vs$}
        \label{variableV}
        
        We can also let $\Vs$ be variable. Instead of being given set values, we are given a feasible set $\mathcal V$. We assume that $\mathcal V\subset\fbraces{\Vs\in\R n:V_i\geq0\;\forall i,\,\sum_{i}^nV_i=1}$ and that $\mathcal V$ is a polytope (expressed by linear constraints).
        To simultaneously find the GATE, subject to $\Vs\in\mathcal V$, that is most easily estimable and the weights $\Ws$ to estimate this GATE, we propose to solve the following optimization problem
        \begin{align}
        \underset{\substack{\Vs \in \mathcal{V}\\ \Ws \in \mathcal{W}}}{\min}\prns{\Delta^2_1(\Ws,\Vs)+\Delta^2_0(\Ws,\Vs)+\frac{\lambda_0}{n^2} \| I_SI_0 \Ws \|_2^2+\frac{\lambda_1}{n^2} \| I_SI_1 \Ws \|_2^2}
        .\label{kom_cmse2}
        \end{align}
        When $\mathcal V=\{V_{1:n}\}$ is a singleton, this optimization problem is the same as that in eq.~\eqref{kom_cmse1}.
        Again, we can show that the optimization problem \eqref{kom_cmse2} reduces to a linearly-constrained convex-quadratic optimization problem:
        \begin{equation}
        \label{qpkom2}
        \begin{aligned}
        &\underset{\substack{\Vs\in\mathcal V,\Ws\geq0,\\\Ws^TI_{S}I_1e_n=n,\\\Ws^TI_0e_n=n}}{\min}
        (\Ws,\Vs)^T \left(
        \frac{1}{n^2} \sum_{t \in \{0,1\}} \begin{bmatrix}
            I_{S}I_{t} K_t I_{t}I_{S} + \lambda_t I_{S}I_t & -I_{S} I_{t}K_t  \\
            -K_t I_{t} I_{S} & K_t 
          \end{bmatrix}
         \right)(\Ws,\Vs).
        \end{aligned}
        \end{equation}
        
        The solution to the optimization problem \eqref{qpkom2} provides both weights $V^*_{1:n}$ that define a GATE of interest and the weights $W^*_{1:n}$ to estimate it. The weights $V_{1:n}$ are chosen in order to allow for minimal CMSE. That is, it focuses on the subpopulation where the average effect on which is easiest to estimate by KOM. We discuss this further in Section~\ref{chooseV}.
        
        When we use $\mathcal V=\fbraces{\Vs\in\R n:V_i\geq0\;\forall i,\,\sum_{i}^nV_i=1}$, we term the resulting GATE estimand Kernel Optimal Weighted ATE (KOWATE).
        We can also construct other causal estimands by choosing different $\mathcal V$. For instance, we may restrict to an unweighted subsample as in the OSATE of \cite{crump2009dealing} by choosing $\mathcal V=\fbraces{\Vs\in \fbraces{ 0,1/n'}^n:\sum_{i}^nV_i=1}$, where $n'$ is a chosen subsample size. 
        We refer to this as Kernel Optimal SATE (KOSATE). Table \ref{table1b} summarizes these causal estimands. It is worth noticing that, other causal estimands can be easily constructed by plugging the set of overlap or truncated weights of \cite{crump2009dealing,li2018balancing} as fixed $\Vs$ in the optimization problem. In the next Section we provide more insight on the set of weights $\Vs$ chosen.
        
         \begin{table}[t!]
        \centering
        \caption{Summary of causal estimands, the corresponding of GATE weights $\mathcal{V}$, and the set of optimal KOM weights $W^{\op{KOM}}_i$.
        \label{table1b}}
        \begin{tabular}{cccc}
        \hline
        Estimand &         $\mathcal{V}$ & Type &        $W^{ \op{KOM} }_i$  
        \\ 
        \hline
        SATE     &  $\braces{S_{1:n}}$  &Fixed ($\abs{\mathcal V}=1$)  &   $W^{*}_i$  from \eqref{kom_cmse1}                       \\
        SATT         & $\braces{\frac{n T_{1:n}}{\sum_{j=1}^n T_j}}$&Fixed ($\abs{\mathcal V}=1$) & $W^{*}_i$  from \eqref{kom_cmse1}  \\
        TATE     &  $\braces{\frac{n(1-S_{1:n})}{\abs{\Unt}}}$&Fixed ($\abs{\mathcal V}=1$) &   $W^{*}_i$  from \eqref{kom_cmse1}                       \\
        OWATE     &  $\braces{n \frac{\phi(X_{1:n})(1-\phi(X_{1:n}))}{n_{\op{O}}}}$&Fixed ($\abs{\mathcal V}=1$) &   $W^{*}_i$  from \eqref{kom_cmse1}                       \\
        KOWATE     & $\fbraces{\Vs\in \mathbb{R}^n_{\geq 0}:\sum_{i}^nV_i=1}$&Variable         &        $W^{*}_i$  from \eqref{kom_cmse2}         \\
        KOSATE      & $\fbraces{\Vs\in \fbraces{0,1/n'}^n:\sum_{i}^nV_i=1}$&Variable    &        $W^{*}_i$  from \eqref{kom_cmse2}        \\
         \hline
        \end{tabular}
        \end{table}

%
%

\subsection{What populations are KOWATE and KOSATE choosing?}
        \label{chooseV}
        In the previous Sections, we have seen that by changing the set of weights $\Vs$, we change the target causal estimand considered and consequently the target population under study. In addition, we have seen that this set of weights can be optimally obtained by letting $\Vs$ be variable and be chosen by the optimization problem \eqref{kom_cmse2}. 
        The idea is to pick the subpopulation that is easiest to estimate by KOM. This subpopulation will emphasize areas with better overlap, where overlap is characterized in terms of worst-case moment discrepancies as defined by the kernels, rather than in terms of (unknown) propensity scores.
        
\begin{figure}[H] 
        \begin{center}
        \includegraphics[scale=.46]{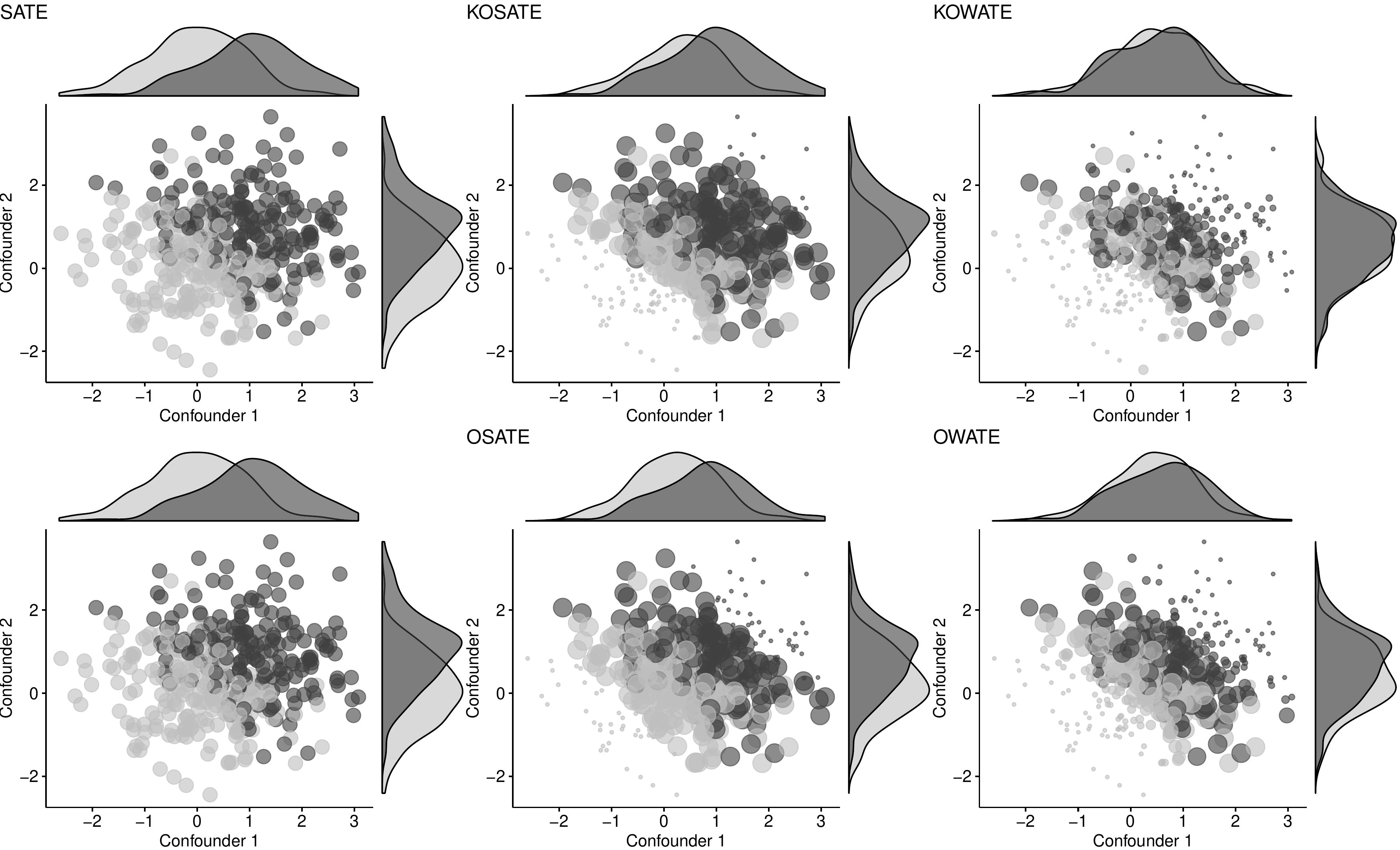}
        \end{center}
        \caption{\footnotesize \textit{Weigths $\Vs$}: Scatterplots between two confounders, confounder 1 in the X-axis and confounder 2 in the Y-axis, weighted by the set of weights $\Vs$, obtained when targeting SATE (first top and bottom panels), KOSATE (second top panel), KOWATE (third top panel), OSATE (second bottom panel) and OWATE (third bottom panel). The histograms on the top and right axes represent the distributions of the confounders across treated (dark-grey) and control (light-grey).  
        \label{figexplainw} }
        \end{figure}

        In this Section, we illustrate this in a simple simulated example described in Figure \ref{figexplainw}. Specifically, Figure \ref{figexplainw} shows scatterplots between two confounders, one on the vertical axis and one on the horizontal axis, weighted by the weights $\Vs$, obtained when targeting SATE (first column of Figure \ref{figexplainw} 
        , KOSATE (second top panel), KOWATE (third top panel), OSATE (second bottom panel) and OWATE (third bottom panel). The histograms on the top and right axes represent the distributions of the confounders across treated (dark-grey) and control (light-grey). 
        The data was generated to exhibit practical positivity violations and we provide more details on the data generation in the simulation section (Section \ref{simu}). 
        
        When targeting SATE, we consider a fixed $\Vs$ that is equal to 1 for all units in the sample. On the other hand, all of KOSATE, OSATE, KOWATE, and OWATE focus on the area of confounders with high overlap.
        In practice, we find that this translate to better performance, as seen in Table \ref{tab2} in our case study.
        KOSATE and OSATE do this while restricting to either including or excluding samples, as can be seen by the two point sizes in Figure \ref{figexplainw}. KOWATE and OWATE consider a range of weights, as can be seen by the variable point sizes.
        Visually, the weights that define the GATE for KOSATE and OSATE are similar as they both focus on the area of overlap; the same for KOWATE and OWATE. 
        The differences are that KOWATE and KOSATE guard against possible misspecification of propensity models and that they target the CMSE of the estimator itself, rather than the asymptotic variance, and therefore they account for the desired precision of the KOM estimate that will be applied. 
        We provide a deeper study of this in Section \ref{more_on_VW} of the Supplementary Material, where we consider the effects of misspecification as well as how the weights $\Ws$ differ as well between the methods.

%
%

\subsection{Consistency}

        In this Section we study the consistency of the proposed weighted estimator with respect to the true causal estimand GATE (for $\Vs$ both fixed and variables). 
        \begin{theorem}
        \label{thm3}
            For $\Vs$ given, let $W_{1:n}^{*}$ be the set of optimal weights obtained by solving optimization problem~\eqref{kom_cmse1}. Suppose Assumptions \ref{unconftrt}--\ref{bound2} hold. Assume also, $\|g\|<\infty$, $\Eb{\mathcal K_t(X_i,X_i)}<\infty$, $\sigma_t\leq\overline\sigma$,  and $\Eb{V_i^2}<\infty$. Then, $\hat{\tau}_{W_{1:n}^{*}} - \tau_V = O_p(1/\sqrt{n})$.
        \end{theorem}
        The above theorem shows that for any GATE estimand, under appropriate assumptions, the KOM estimate is root-$n$ consistent.
        
        The assumption about the kernel can be automatically satisifed by using a bounded kernel, such as the Gaussian or Matern kernels.
        The assumption of $\|g\|<\infty$ requires no model misspecification. We can relax this assumption if we use a universal kernel, such as Gaussian, but the rate may deteriorate from $O_p(1/\sqrt{n})$ to $o_p(1)$ as we need to include a vanishing approximation term. For brevity, we omit the details.
        
        To apply Theorem~\ref{thm3} to the case of variable $\Vs$, note that the solution $\Vs^*$ of problem~\eqref{kom_cmse2} is a function of $H_{1:n}$ and is therefore honest (satisfies Assumption~\ref{honest}) and that, given these $\Vs=\Vs^*$, the solution $\Ws^*$ to problem~\eqref{kom_cmse1} is the exactly same as that in problem~\eqref{kom_cmse2}, as it can just be viewed as a nested minimization problem (once in $\Vs$ and once in $\Ws$). So to apply Theorem~\ref{thm3} to the case of variable weights, we need only guarantee that $\E[V_i^2]<\infty$. We can either take that as an assumption, or we can enforce it in the construction of $\mathcal V$ by including a bound. In practice, we find that this is not necessary.
        
        
        
        
%
%

\subsection{Kernel choice, automatic selection of the its hyperparameters, and other practical guidelines}
\label{guidelines}

        Solutions to the optimization problems \eqref{kom_cmse1} and  \eqref{kom_cmse2} depend on the choice of the kernel and its hyperparameters. In this Section we provide some practical suggestions on how to chose them. Generally, we suggest the use of a polynomial Mahalanobis kernel:
        \begin{equation}\label{polykernel}
        \mathcal K_t(x,x')=\gamma_t(1+\theta_t(x-\hat\mu_n)^T\hat\Sigma_n^{-1}(x'-\hat\mu_n))^d,
        \end{equation}
        
        \noindent
        where $\hat \mu_n$ is the sample mean, $\hat\Sigma_n$ is the sample covariance, $d$ is the parameter that controls the degree of the polynomial, $\theta_t$ is a parameter that controls the importance of higher orders degrees, and $\gamma_t$ controls the overall scale of the kernel.   To avoid unit dependence and confounders with high variance dominating those with smaller ones, by proposing the aforementioned kernel, we are suggesting to normalize confounders to have mean 0 and variance 1. Based on the results of the next Section and previous simulation studies \citep{kallus2018more}, we suggest using $d\geq2$. Alternatively, we may additionally replace $\hat\Sigma_n^{-1}$ with a matrix parameter to be tuned. Alternative choices for the kernel include Gaussian or Matern. These have the benefit of being universal approximators, but practically we find a polynomial kernel is sufficient.
        
        To tune the kernel's hyperparameters, $\theta_t$ and $\gamma_t$, we propose to use marginal likelihood, a well-known model selection criteria for Gaussian processes \citep{rasmussen2010gaussian}. To do so, we specify two Gaussian process priors, $f_1$ and $f_0$ with covariances specified by the kernels $\mathcal{K}_1$ and $\mathcal{K}_0$. We suppose that the the potential outcome $Y_i(t)$ was observed from $f_{t}(X_i)$ with Gaussian noise of variance $\sigma^2_t$. We then maximize the marginal likelihood (marginalizing over the Gaussian process) of seeing the data with respect to the hyperparameters, $\theta_t$, $\gamma_t$, and $\sigma^2_t$.
        
        In addition, optimization problems \eqref{kom_cmse1} and  \eqref{kom_cmse2} depend on the choice of the hyperparameters $\lambdas$, which control the trade-off between bias and variance. In order to target the total error (\ie, the CMSE) we set $\lambda_t = \sigma^2_t / \gamma^2_t$ for $t \in \lbrace 0,1 \rbrace$, in hopes to target the worst-case CMSE. 
         When using this estimated CMSE-optimal $\lambdas$, we are targeting the estimator with minimal total error. 
         When $\lambdas$ equals zero, we instead target minimal bias. As a general practical advice, we suggest using the estimated optimal $\lambdas$ obtained by using plug-in estimates from GPML as aforementioned. 
        
        Several software packages implementing marginal likelihood can be used to tune hyperparameters and a variety of solvers can be used to solve linearly-constrained convex-quadratic optimization problems. We suggest using the \textsf{GaussianProcessRegressor} package from \textsf{scikit-learn} \citep{scikit-learn} for tuning the hyperparameters and \textsf{Gurobi} \citep{optimization2014inc_2} for solving quadratic optimization problems. 
        In practice, \textsf{Gurobi} sometimes fails due to the quadratic objective being numerically non-PSD (despite being PSD in theory). We have found this occurs sometimes when using a high-degree kernel and variable $\Vs$. We found that this is easily fixed without materially changing the results by adding $10^{-8}I$ to the quadratic objective matrix to inflate its spectrum slightly.
        

         We suggest to use Wald confidence intervals together with the robust ``sandwich'' standard error estimator as previously suggested by other authors \citep{hernan2001marginal,robins2000marginal,freedman2006so,imai2014covariate}. We provide more details on standard error estimation and coverage in Section \ref{se_cove} of the Supplementary Material.


%
%

\section{Simulations}
\label{simu}

        In this Section, we present the results of a simulation study aimed at comparing KOM with IPW, overlap weights, truncated IPW, and outcome regression modeling in estimating GATE with respect of absolute bias and root MSE, across levels of practical positivity violations and across levels of misspecification. In summary, KOM showed a consistently low absolute bias and root mean squared error (RMSE) across all of the considered scenarios.

%
%

\subsection{Setup}
        \label{setup}

        We considered a sample size of $n=400$. We computed the potential outcomes $Y(1), Y(0)$ from the following models: $Y_i(0) =  3( X_{i,1} + X_{i,2}) + N(0,1)$, $Y_i(1) = Y_i(0) + \delta$, where $\delta=4$. We computed the observed outcome as $Y_i = Y_i(0) (1-T_i) + Y_i(1) T_i$, where $T_i \sim \text{binom}(\pi_i)$, $\pi_i = (1+\exp{(-\alpha(-1.5 + 1.5X_{i,1} +  1.5X_{i,2})))^{-1}}$,  $X_{k,i} \sim \text{N}(0.5,1)$, $k=1,2$.  The true GATE was computed as $\tau_V=\sum_{i=1}^nV_i(Y_i(1)-Y_i(0))$. We considered $S_i=1$ for all units in the sample.
        We consider several $V_i$, namely, (a) KOWATE; (b) KOSATE where we set $n'=n_T$ equal to the number of units chosen by OSATE, (c) SATE, (d) OWATE, (e) OSATE with $\alpha=0.1$. 
        For KOWATE and KOSATE we consider the KOM weights given by problem \eqref{kom_cmse2}.
        For SATE we consider several estimates: (a) KOM as in the optimization problem \eqref{kom_cmse1}; (b) IPW; and (c) using outcome regression modeling. For OWATE and OSATE, we use the estimated propensity 
        to define $\Vs$.
        %
        %
        To estimate OSATE, we computed the set of truncated weights setting $\alpha=0.1$. We used a product of polynomial-degree-2 kernels for KOM. We modeled the propensity score, $\phi(X_i)$, by using a polynomial-degree-4 logistic regression and we used a polynomial-degree-4 regression model for outcome regression modeling. 
        We evaluated the performance of the proposed method across levels of practical positivity violation and misspecification as described in Section \ref{overlap} and Section \ref{miss} respectively. When the solver failed to solve the optimization problem \eqref{kom_cmse1} or \eqref{kom_cmse2} when using a product of polynomial-degree-2 kernels, we rerun the same optimization problem by considering a KOM polynomial degree 3. If also KOM polynomial-degree-3 failed then we considered KOM polynomial-degree-2. We estimated the estimand of interest by plugging in the set of obtained weights into the weighted estimator $\hat \tau_{W}$. We used \textsf{scikit-learn} (through the \textsf{R} package \textsf{reticulate}) to tune the hyperparametes and the \textsf{R} interface of \textsf{Gurobi} to obtain the set of KOM weights.

%
%

\subsubsection{Estimating GATE across levels of practical positivity violation}
\label{overlap}

        To evaluate the performance of the proposed methodology across levels of practical  positivity violation, we let $\alpha$ vary between $0.1$ and $1$.  We considered 10 levels.  We refer to $\alpha=0.1$ as weak practical positivity violation, to  $\alpha=0.5$ as moderate, and to $\alpha=1$ as strong. In our simulation scenario the propensity score, $\phi(X_i)$, ranged between 0.37 and 0.63 under weak violation, between 0.07 and 0.92 under moderate violation and between 0.007 and 0.993 under strong violation (average of min/max of estimated propensities over simulations under no misspecification).

\subsubsection{Estimating GATE across levels of misspecification}
\label{miss}

    We also evaluated the performance of the proposed methodology across levels of misspecification.  To do so, we generated $Z_1 = X_2/\exp{(X_1)}$ and $Z_2 = \log(|X_2|)$ and considered a convex combination between the correct variables $(X_1,X_2)$ and the misspecified variables $(Z_1,Z_2)$, $X_1 = \gamma X_1  + (1-\gamma)Z_1$ and $X_2 = \gamma X_2  + (1-\gamma)Z_2$. 
    We considered 3 levels, $\gamma=1$ which we refer to as correct specification (which is also overparametrized because we use the polynomial models previously described in all scenarios), $\gamma=0.5$ which we refer to as moderate misspecification, and $\gamma=0$ which we refer to as strong misspecification. 

%
%

\begin{figure}[H] 
        \begin{center}
        \includegraphics[scale=.6]{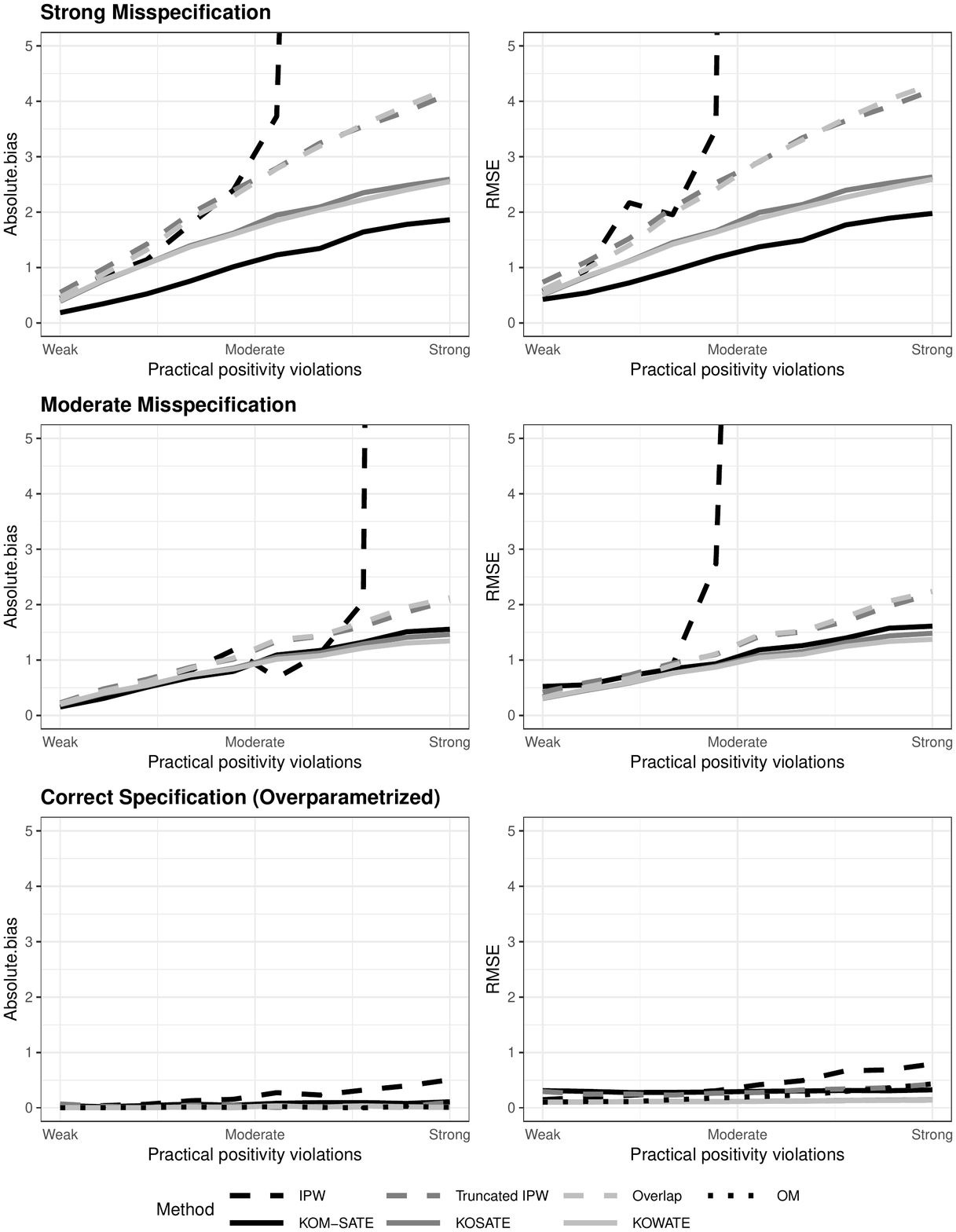}
        \end{center}
        \caption{\footnotesize (Estimated optimal $\lambdas$): Absolute bias (left panels) and RMSE (right panels) of SATE estimated by using KOM (solid-black)(which we refer to as KOM-SATE), KOSATE  by using KOM (solid-dark-grey), KOWATE estimated by using KOM (solid-light-grey), SATE estimated by using IPW (long-dashed-black), OSATE estimated by using truncated weights (long-dashed-dark-grey), OWATE estimated by using overlap weights (long-dashed-light-grey), and SATE estimated by using outcome regression modeling (dotted-black)(which we refer to as OM) when increasing the strength of practical positivity violation under strong misspecification (top panels), moderate misspecification (middle panels) and correct (and overparametrized) specificiation (bottom panels), $n=400$.  
        \label{figowate1} }
        \end{figure}

\subsection{Results}

        In this Section we discuss the results of our simulation study. In summary, KOM outperformed IPW, overlap, truncated weights and outcome regression modeling with respect of absolute bias and RMSE in estimating GATE across levels of practical positivity violation under both moderate and strong misspecification. 
        
        \subsubsection{Results across levels of practical positivity violations and model misspecification}
        \cite{kallus2018more} presented KOM for SATE. The Authors showed that KOM outperformed IPW, truncated IPW, propensity score matching, regression adjustment, CBPS and SBW with respect to bias and MSE across most of the considered levels of practical positivity violation and considered scenarios.  In addition, the authors showed that KOM for SATE outperformed the other methods especially under strong practical positivity violation. Figure \ref{figowate1} shows the absolute bias (left panels) and RMSE (right panels) of SATE estimated by using KOM (KOM-SATE; solid-black), KOSATE  by using KOM (solid-dark-grey), KOWATE estimated by using KOM (solid-light-grey), SATE estimated by using IPW (long-dashed-black), OSATE estimated by using truncated weights (long-dashed-dark-grey), OWATE estimated by using overlap weights (long-dashed-light-grey), and SATE estimated by using outcome regression modeling (OM; dotted-black), with estimated optimal $\lambdas$ (\ie, $\lambda_t=\frac{\sigma_t^2}{\gamma_t^2}$). The top panels of Figure \ref{figowate1} show absolute bias and RMSE across levels of practical positivity violations under strong misspecification, while the middle and bottom panels under moderate misspecification and correct specification, respectively. In summary, KOM showed a consistently low absolute bias and RMSE across all the considered scenarios, outperforming the other methods especially under strong practical positivity violation and strong misspecification (top-right panel). KOM matched the performance of the other methods with respect of absolute bias and RMSE under weak levels of practical positivity violations and correct model misspecification. We obtained similar results when $\lambdas=0$ (Figure \ref{figowate0} in the Supplementary Material). IPW exhibited extremely high absolute bias and RMSE under moderate to strong practical positivity violations and moderate to strong model misspecification. Under moderate and strong misspecification, OM resulted in even higher absolute bias and RMSE across all levels of practical positivity violations that the results are outside the plot region in the top and middle panels of Figures \ref{figowate1} and \ref{figowate0}.  We provide additional simulations results in Section \ref{supmat_simu} in the Supplementary Material.

%
%

\section{Application Case Studies}
\label{illustrations}

        In this Section we present two empirical applications of the proposed methodology. In the first, we apply KOM in the evaluation of two spine surgical interventions, laminectomy alone versus fusion-plus-laminectomy, on the Oswestry Disability Index (ODI), among patients  with  lumbar  stenosis  or  lumbar  spondylolisthesis. In the second, we apply KOM in the evaluation of peer support on CD4 cell count at 12 months after trial recruitment among patients affected by HIV, in two target populations where patients were healthier compared to those of the trial population.

%
%

\subsection{The effect of fusion-plus-laminectomy on ODI}

        In this Section we apply KOM in the evaluation of two spine surgical interventions, laminectomy alone versus fusion-plus-laminectomy, on the Oswestry Disability Index (ODI), among patients  with  lumbar  stenosis  or  lumbar  spondylolisthesis. Briefly, lumbar stenosis is caused by the narrowing of the space around the spinal cord in the lumbar spine \citep{resnick2014guideline}. Lumbar  spondylolisthesis is caused by the slippage of one vertebra on another. These pathologies lead to low back and leg pain, ultimately limiting the quality of life of those patients affected by them \citep{waterman2012low}. In case these pathologies are not anymore controlled by medications or physical therapy, surgical interventions may be needed. Typically, patients with lumbar stenosis are treated with laminectomy alone while those with lumbar spondylolistheses with fusion-plus-laminectomy  \citep{resnick2014guideline, eck2014guideline,raad2018trends}.  In addition, laminectomy alone is done to patients with leg pain, while fusion-plus-laminectomy to patients with mechanical back pain \citep{resnick2014guideline}. This surgical practice leads to a practical positivity violation.  
        
        Differently from other medical areas where randomized controlled trials are the gold standard to evaluate interventions, the use of randomized controlled trials to evaluate surgical interventions is rare. This is due to practical and methodological issues \citep{carey1999randomized}. Lately, a number of large real-world observational datasets have collected information about surgical interventions and outcomes. However, these datasets are purely observational and confounding must be carefully taken into account. Furthermore, the assumption of correct model specification is hardly ever met. To overcome these challenges, in this Section we evaluate the effect of fusion-plus-laminectomy on ODI by estimating SATE, KOWATE, and KOSATE using KOM.

\subsubsection{Study population}

        We used data from a single-institutional subset of the Spine QOD registry \citep{qod}. QOD was launched in 2012 with the goal of evaluating the effectiveness of spine surgery interventions on the improvement quality of life, pain, and disability. The registry contains clinical and demographic information as well as patient-reported outcomes. We restrict our study to patients who had the their first spine surgery intervention, \ie,~primary surgery. Demographic and clinical information was collected at the time of the patient interview which happened before surgical intervention. The outcome under study, ODI, was collected at 3-month follow-up. The study subset was composed of 313 patients. Two-hundred forty-nine (79\%) received laminectomy alone and 64 (21\%) fusion-plus-laminectomy. We identified as potential confounders the following variables: biological sex (female vs. male),  lumbar stenosis (yes vs. no), lumbar spondylolistheses (yes vs. no), back pain (score from 0 to 10), leg pain (score from 0 to 10), and activity at home (yes vs. no), activity outside home (yes vs. no). As previously described, spine surgical practice may lead to a practical violation of the positivity assumption.  For example, in our subset, less then 1$\%$ of patients with low-to-moderate leg pain were treated with fusion-plus-laminectomy.

\subsubsection{Models setup}

        We estimate SATE by solving optimization problem \eqref{kom_cmse1} with $\Vs=e$, and KOWATE and KOSATE by solving optimization problem \eqref{kom_cmse2} where we set $\fbraces{\Vs\in \mathbb{R}^n_{\geq 0}:\sum_{i}^nV_i=1}$ and $\fbraces{\Vs\in \lbrace 0,1 \rbrace :\sum_{i}^nV_i=n_{\op{T}}}$, respectively. We obtained $n_{\op{T}}$ by summing truncated weights obtained by using a logistic regression model and setting $\alpha=0.1$. Once the set of weights were obtained, we plugged them into a weighted ordinary least squares estimator.  We used \textsf{scikit-learn} (through the \textsf{R} package \textsf{reticulate}) to tune the hyperparametes and the \textsf{R} interface of \textsf{Gurobi} to obtain the set of KOM weights. We computed robust (sandwich) standard errors in each case. We used the \textsf{R} packages \textsf{lm} for estimating SATE, KOWATE and KOSATE and \textsf{sandwich} to estimate robust standard errors.

\subsubsection{Results}

\begin{figure}[H] 
        \begin{center}
        \includegraphics[scale=.55]{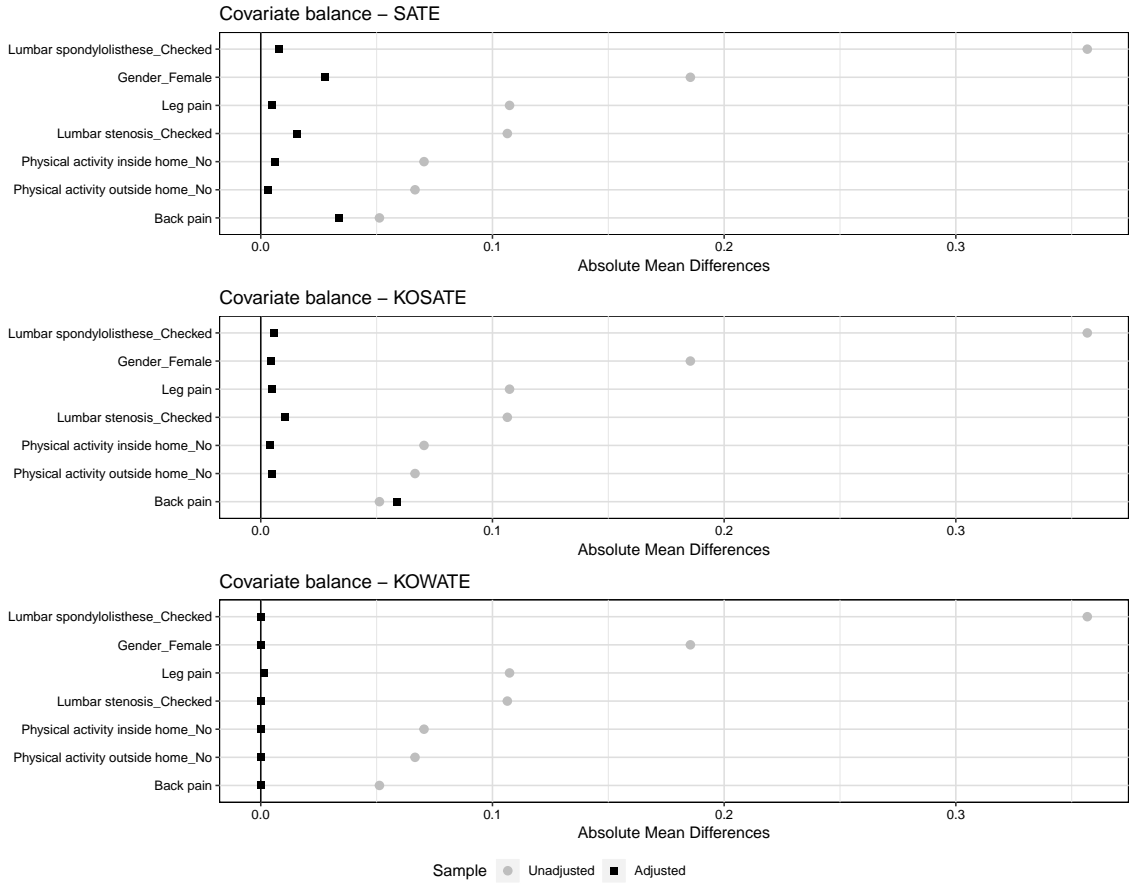}
        \end{center}
        \caption{\footnotesize Covariate balance with respect to SATE (top panel), KOSATE (middle panel) and KOWATE (lower panel). The black dots reflect the level of balance after weighting for SATE, KOSATE and KOWATE weights, while the light-grey dots show the unadjusted balance.
        \label{fig_covbal} }
        \end{figure}

        In this Section we present the results of our analysis. Previous randomized trials showed no statistically significant difference between laminectomy alone versus fusion-plus-laminectomy on ODI \citep{forsth2016randomized,ghogawala2016laminectomy}. The proposed methodology consistently showed similar results to those of \cite{forsth2016randomized,ghogawala2016laminectomy}. Specifically, Table~\ref{tab2} shows point estimates and standard errors with respect to SATE, KOWATE, and KOSATE. While the unadjusted method, \ie, naive method regressing only the treatment on the outcome, shows a significant effect of fusion-plus-laminectomy on ODI, adjusted estimates from SATE, KOWATE and KOSATE show a non statistically significant effect of it. Standard errors are lower for KOWATE and KOSATE compared to SATE. Figure \ref{fig_covbal} shows the covariate balance with respect to SATE (top panel), KOSATE (middle panel) and KOWATE (lower panel). The black dots show the level of balance after weighting, while the light-grey dots show the unadjusted balance. KOWATE provides the lowest covariate balance compared with SATE and KOSATE.  Finally,  based on the results obtained by applying KOM, we conclude that fusion-plus-laminectomy has no statistically significant effect on ODI.

        \begin{table}[H]
        \centering
        \caption{The effect of fusion-plus-laminectomy on ODI \label{tab2}}\setlength{\tabcolsep}{1em}
        \begin{tabular}{lcccc}
          \hline
           & SATE & KOSATE & KOWATE & Unadjusted \\ 
          \hline
          $\hat\tau_W$ $(SE)$ & 1.33 (3.98) & 2.54 (2.56) & 3.03 (2.38)  & 5.09* (2.31)\\ 
           \hline
           \multicolumn{4}{c}{\footnotesize * indicates statistical significance at the 0.05 level.}
        \end{tabular}
        \end{table}

%
%

\subsection{Evaluating the impact of peer support on CD4 cell count in target populations of healthier people}

        In the past four decades, the HIV epidemic has become a global epidemic 
        with major social and economical consequences. People living with HIV (PLWH) have benefited from the services of peer support, such as peer-to-peer counseling, support groups  and  home-based
        adherence counseling. Although several studies have shown the positive effects of peer support on quality of life and on coping with stigma and discrimination 
        , few studies have evaluated its effect on treatment outcomes such as CD4 cell count at 12 month after peer support initiation. The CD4 cell count provides an indication of the health of the immune system of a PLWH. Normal ranges are between 500 cells/mm$^3$ and 1,500 cells/mm$^3$, and it decreases when a person is infected by HIV, leading to AIDS when below 200 cells/mm$^3$. In a recent randomized trial, \cite{sonnerborg2016impact} showed that peer support, defined as home-based adherence counseling, did not have an effect on CD4 cell count at 12 months and other treatment outcomes, such as virological failure. The study was conducted in a resource-limited settings, in which the majority of the PLWH in the sample were severely immune-suppressed with low CD4 cell count at baseline. We are interested in answering the question: what is the impact of peer support in a target population in which PLWH have a higher CD4 cell count at baseline, \ie, they are healthier at the onset.
        To do so, in this Section, we use outcome data from \cite{sonnerborg2016impact} and apply KOM to evaluate the impact of peer support on CD4 cell count in the general population provided by the Multicenter AIDS Cohort Study \citep{MACS}, a prospective observational study collecting demographic and economic information of the HIV-infection. Specifically, we considered two target populations, (1) those PLWH in the MACS dataset that received treatment after 2001 (which we refer to MACS-1), and (2) those that received treatment after 2010 (which we refer to MACS-2). These two populations reflect the fact that in more recent years, PLWH get detected earlier, leading to healthier PLWH at baseline. 

\subsubsection{Study and target populations}

        We restrict our analysis to the subset of 492 PLWH with complete information about CD4 cell count at baseline, age and CD4 cell count at 12 months. We considered CD4 cell count and age at baseline as possible confounders. PLWH in the trial had an average CD4 of 120 cells/mm$^3$ with more than 80\% diagnosed with AIDS at the time of trial initiation. PLWH in the target population MACS-1 had an average CD4 cell count at baseline of 450 cells/mm$^3$, while those in MACS-2 of 640 cells/mm$^3$, again suggesting that PLWH in the two target populations considered are healthier. The mean age in the trial was 32, while that in MACS-1 and MACS-2 was 38 and 36, respectively. The sample sizes of MACS-1 and MACS-2 were 860 and 243, respectively.

\subsubsection{Model setup}
        To estimate TATE, we combined the data in the trial with those in the target population. We set the indicator of sample assignment, $S_i$ equal to 1 when unit $i$-th belonged to the trial and 0 otherwise, and set $V_i=\frac{n(1-S_i)}{\abs{\mathcal{U}}}$. We solved the optimization problem \eqref{qpkom1} to estimate TATE in the two target populations.  Once we obtained the set of KOM weights, we plugged them into a weighted ordinary least squares estimator.  We computed robust (sandwich) standard errors in each case.  We used \textsf{scikit-learn} (through the \textsf{R} package \textsf{reticulate}) to tune the hyperparametes and the \textsf{R} interface of \textsf{Gurobi} to obtain the set of KOM weights, \textsf{lm} for estimating TATE and \textsf{sandwich} to estimate robust standard errors.

\subsubsection{Results}

        Table \ref{tab3} shows the result with respect to TATE across the two target populations MACS-1 and MACS-2. In summary, similarly to as the results in the trial, our results showed a non statistically significant effect of peer support in populations with higher CD4 cell count at onset. We therefore conclude that the results in \cite{sonnerborg2016impact} may be generalized to populations where PLWH had a higher CD4 cell count at baseline. 
        
        \begin{table}[H]
        \centering
        \caption{The impact of peer support on CD4 cell count at 12 months after baseline in the original trial of \cite{sonnerborg2016impact} (SATE) and in the two target populations with higher CD4 cell count at onset (TATE). \label{tab3}}
        \setlength{\tabcolsep}{1em}
        \begin{tabular}{lccc}
          \hline
         $\hat\tau_W$ $(SE)$ & SATE & TATE  \\ 
          \hline
          Trial & -1.3 (16.1) & $-$  \\
          MACS-1  & $-$ & -16.9 (60.6) \\ 
          MACS-2  & $-$ & -28.6 (44.6)  \\ 
           \hline
        \end{tabular}
        \end{table}

%
%

\section{Conclusion}
\label{conclusions}
        In this paper we presented a general causal estimand, GATE, that unified previously proposed causal estimand, such as SATE, OWATE, OSATE and TATE among others and motivated the formulation of new ones. We also presented and applied KOM to optimally estimate GATE. KOM directly and optimally control both bias and variance which leads to a successful mitigation of possible model misspecifications while controlling precision. In addition, by easily modifying the optimization problem that is fed to an off-the-shelf solver, the proposed method effectively target different causal estimands of interest. Furthermore, by automatically learning the structure of the data, KOM allows to balance linear, nonlinear, additive, and non-additive covariate relationships. One future direction may be to extend KOM for GATE in the longitudinal setting with time-dependent confounders, extending the work of \cite{2018arXiv180601083K} to more general estimands.

\bibliographystyle{chicago}
\bibliography{gate}

\newpage
\begin{center}
{\large\bf SUPPLEMENTARY MATERIAL}
\end{center}

\begin{description}

\item[R-code:] R-code containing code to perform the simulations and the analyses of the case-studies described in the article. 

\item[Additional results:] Introduction: Related work; KOM for GATE: Figures \ref{figexpwcore}, \ref{figexpvmiss}, \ref{figexpwmiss}; Simulations: Figures \ref{figowate0}, \ref{figowaten}, \ref{figowate_var1}, \ref{figowate_var0}, \ref{figowate_cove1}, \ref{figowate_cove0} and Table \ref{tab_ct_owate}. 

\item[Proofs:] Proofs of Lemma \ref{lemma1}, Theorem \ref{thm1}, \ref{thm2} and \ref{thm3}.

\end{description}

\newpage

\section{Introduction - Additional results}

We provide additional information about the literature on average treatment effects estimation. 

\subsection{Related work}
\label{related_work}

Randomized controlled trials provide unbiased estimates of the SATE. However, their results may not be extended to different populations because trial participants are not representative of the target populations or subgroup of interest. For example, the Women's Health Initiative (WHI) trial \citep{writing2002risks}, contrarily to previous observational studies, found an harmful effect of hormone-therapy on stroke. These discrepancies have been in part attributed to possible differences in the distributions of age and weight between the WHI trial and the real-world practice. Specifically, women in the WHI trial were older than the typical age at which the hormone therapy is taken and were also more obese, leading to an increased risk of stroke \cite{keiding2016perils}. Observational datasets, such as electronic medical registries, are more representative of the real-world clinical practice, and may provide more generalizable results compared to those of the trials. However, despite their potential, these datasets are observational and non-experimental, where the true causal effect is hidden by confounding factors. Thus, while trials provide unbiased estimates of the SATE in populations that are not generalizable as those in an observational dataset, the estimation of causal effects with observational data is hampered by confounding.

Various statistical methods have been proposed in an attempt to control for confounding in observational studies and to generalized trial results to target populations. Among others, methods based on IPW have been widely used. To control for confounding, IPW weights each subject under study by the inverse of the probability of being treated given covariates, \ie~ the propensity score \citep{rosenbaum1983central}, thus mimicking a random treatment assignment as in a trial. In other words, IPW creates an hypothetical population in which covariates are balanced and confounding is consequently removed. IPW has also been used to generalized trial results to target populations of interest. For instance, based upon the work of \cite{cole2010generalizing} and \cite{stuart2011use}, \cite{buchanan2018generalizing} proposed an IPW estimator that weights each trial participants by the inverse of the probability of trial participation conditional on covariates. Outcome regression modeling, where the outcome given covariates is modeled by using a linear or nonlinear regression model have also been used. Specifically, to control for confounding, for each treatment arm, an outcome model is  postulated, and used to predict the outcome. If the chosen model is the true model that generated the outcome, then, it could be used to both control for confounding and evaluate causal effects in populations different than the one in which the model was trained.

Despite their wide applicability, these two methods are both highly sensitive to model misspecification. Specifically, IPW-based methods are sensitive to misspecification of the treatment assignment model, used to construct the weights, while outcome regression modeling to that of the outcome model. In addition, the use of IPW-based methods is jeopardized by their dependence on the positivity assumption \citep{rosenbaum1983central}, which requires that the propensity scores are neither 0 nor 1. Although positivity may hold theoretically, it can be practically violated \citep{petersen2012diagnosing}, \ie, lack of overlap in the covariate distributions between
treatment groups, yielding to propensity scores close to 0 or 1. Practical violations of the positivity assumption leads to extreme weights and erroneous inferences. 

Methods have been proposed to overcome the issue of misspecification. \cite{robins1994estimation} proposed augmented IPW estimators, which combine IPW and outcome models in one doubly robust estimator. Here, doubly robust refers to the fact that an unbiased estimate of SATE can be obtained whenever either of the outcome model or the treatment assignment model is correctly specified, thus being robust to misspecification. However, as noted by \cite{kang2007demystifying}, these methods also suffer from practical violations of the positivity assumption and they are highly biased in case of misspecification of both treatment and outcome models. \cite{imai2015robust} proposed to use the Covariate Balance  Propensity  Score  (CBPS), which finds the logistic regression model that balances covariates. 

Several methods have been proposed in the past decades to overcome the issue of extreme weights in IPW. \cite{robins2000marginal} suggested the use of stabilized inverse probability weights, which are obtained by normalizing the weights by the marginal  probability  of  treatment. \cite{santacatterina2017optimal} proposed to use shrinkage to better control the bias-variance trade-off. \cite{zubizarreta2015stable} proposed Stable Balancing
Weights (SBW), which are the set of weights of minimal sample variance that satisfy a list of approximate moment matching conditions to a level of balance specified by the research.  \cite{cole2008constructing,xiao2013comparison} suggested truncation,  which consists of replacing outlying weights with less extreme ones. A number of approaches have been proposed with the goal of defining a study population in which the average treatment effect can be well estimated while being as inclusive as possible, \ie, defining a study population that has enough overlap between treatment groups. \cite{crump2009dealing} proposed to choose the study population that optimize the variance of the estimated average treatment effect on the study population. The Authors showed that this study population is composed by those units whose propensity scores lie in an interval $[\alpha,1-\alpha]$. They suggest to set $\alpha=0.1$. As previously mentioned, the authors refer to this causal estimand as OSATE. \cite{visconti2018handling,zubizarreta2014matching} proposed to use cardinality matching, in which integer programming is used to find the largest sample with bounded balance. Other methods have been also proposed \citep[among others]{traskin2011defining,king2017balance}. Finally, \cite{crump2009dealing,li2018balancing} proposed to use overlap weights, where each unit’s weight is proportional to the probability of that unit being assigned to the opposite treatment group. As previously mentioned, the authors refer to this causal estimand as OWATE.

The literature of causal estimation is extensive and many other methods have been developed in the recent years \citep[among others]{hirshberg2019minimax,hirshberg2017augmented,hainmueller2012entropy,zhao2017entropy,zhao2019covariate,wong2017kernel}

Our main contributions to this literature are 1) propose a general causal estimand that unifies several causal estimands and motivates the formulation of new ones, 2) present and apply KOM, which by optimally controlling bias and variance, it mitigates possible model misspecification while controlling for precision. 

\section{Kernel Optimal Matching for estimating GATE - Additional results}
\label{more_on_VW}

\begin{figure}[H] 
\begin{center}
\includegraphics[scale=.46]{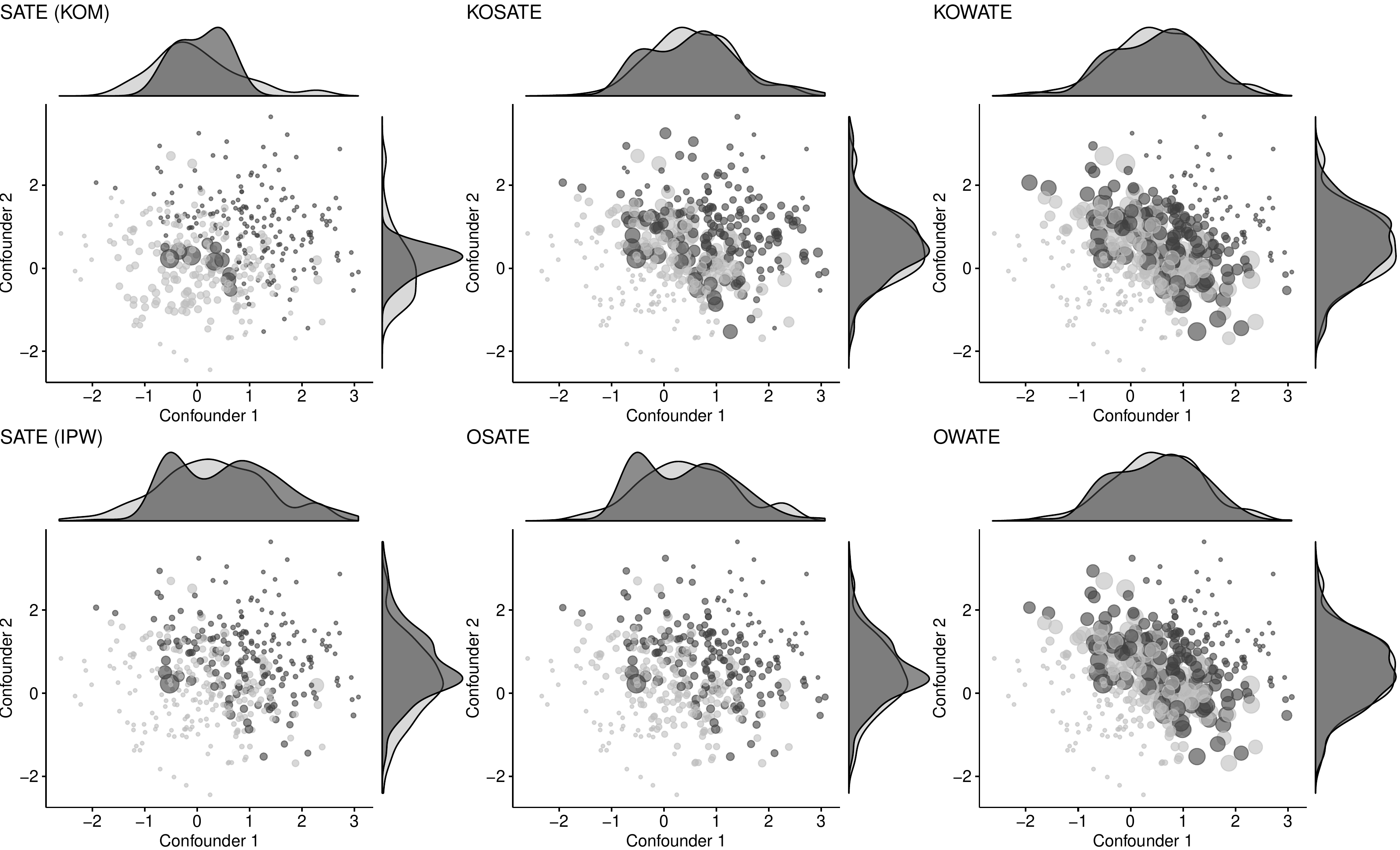}
\end{center}
\caption{\footnotesize \textit{Correct specification - Weights $\Ws$}: Scatterplots between two confounders, confounder 1 in the X-axis and confounder 2 in the Y-axis, weighted by the set of weights $\Ws$, obtained when targeting SATE (first top and bottom panels of Figure \ref{figexplainw}), KOSATE (second top panel), KOWATE (third top panel), OSATE (second bottom panel) and OWATE (third bottom panel) under correct specification. The histograms on the top and right axes represent the distributions of the confounders across treated (dark-grey) and control (light-grey).  
\label{figexpwcore} }
\end{figure}

\begin{figure}[H] 
\begin{center}
\includegraphics[scale=.46]{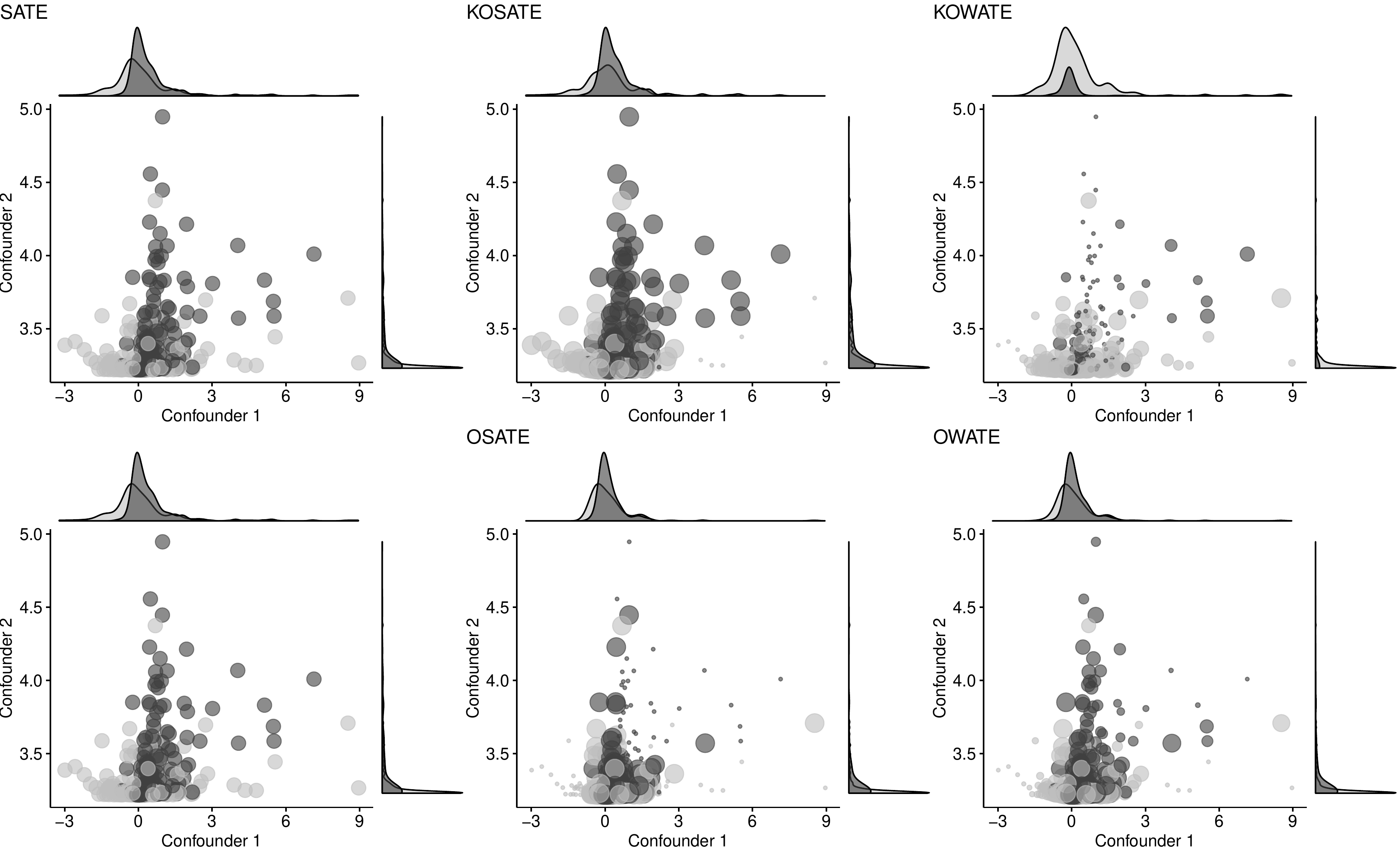}
\end{center}
\caption{\footnotesize \textit{Strong misspecification - Weights $\Vs$}: Scatterplots between two confounders, confounder 1 in the X-axis and confounder 2 in the Y-axis, weighted by the set of weights $\Vs$, obtained when targeting SATE (first top and bottom panels of Figure \ref{figexplainw}), KOSATE (second top panel), KOWATE (third top panel), OSATE (second bottom panel) and OWATE (third bottom panel) under strong misspecification. The histograms on the top and right axes represent the distributions of the confounders across treated (dark-grey) and control (light-grey).  
\label{figexpvmiss} }
\end{figure}

\begin{figure}[H] 
\begin{center}
\includegraphics[scale=.46]{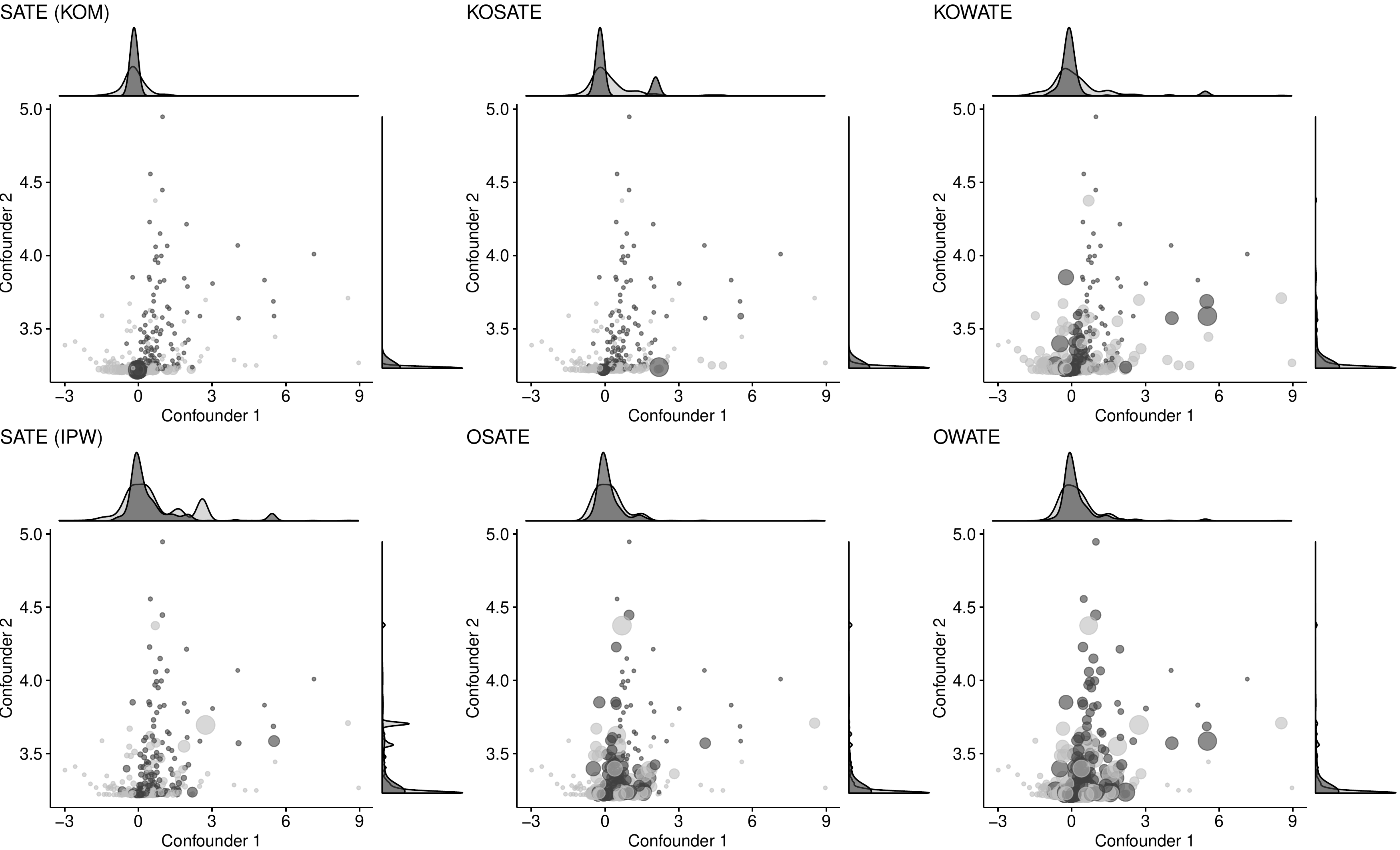}
\end{center}
\caption{\footnotesize \textit{Strong Misspecification - Weights $\Ws$}: Scatterplots between two confounders, confounder 1 in the X-axis and confounder 2 in the Y-axis, weighted by the set of weights $\Ws$, obtained when targeting SATE (first top and bottom panels of Figure \ref{figexplainw}), KOSATE (second top panel), KOWATE (third top panel), OSATE (second bottom panel) and OWATE (third bottom panel) under strong misspecification. The histograms on the top and right axes represent the distributions of the confounders across treated (dark-grey) and control (light-grey).   
\label{figexpwmiss} }
\end{figure}

In this Section, we provide additional results related to Section \ref{chooseV}. Specifically, in Figure \ref{figexpwcore} we show the scatterplots between the two confounders weighted by the weights $\Ws$ under correct specification as described in Section \ref{simu}. Figures \ref{figexpvmiss} and \ref{figexpwmiss} show scatterplots weighted by $\Vs$ and $\Ws$ respectively, under strong misspecification. Weights are standardized to mean one for comparison.

\section{Simulations - Additional results}
\label{supmat_simu}

\begin{figure}[H] 
\begin{center}
\includegraphics[scale=.6]{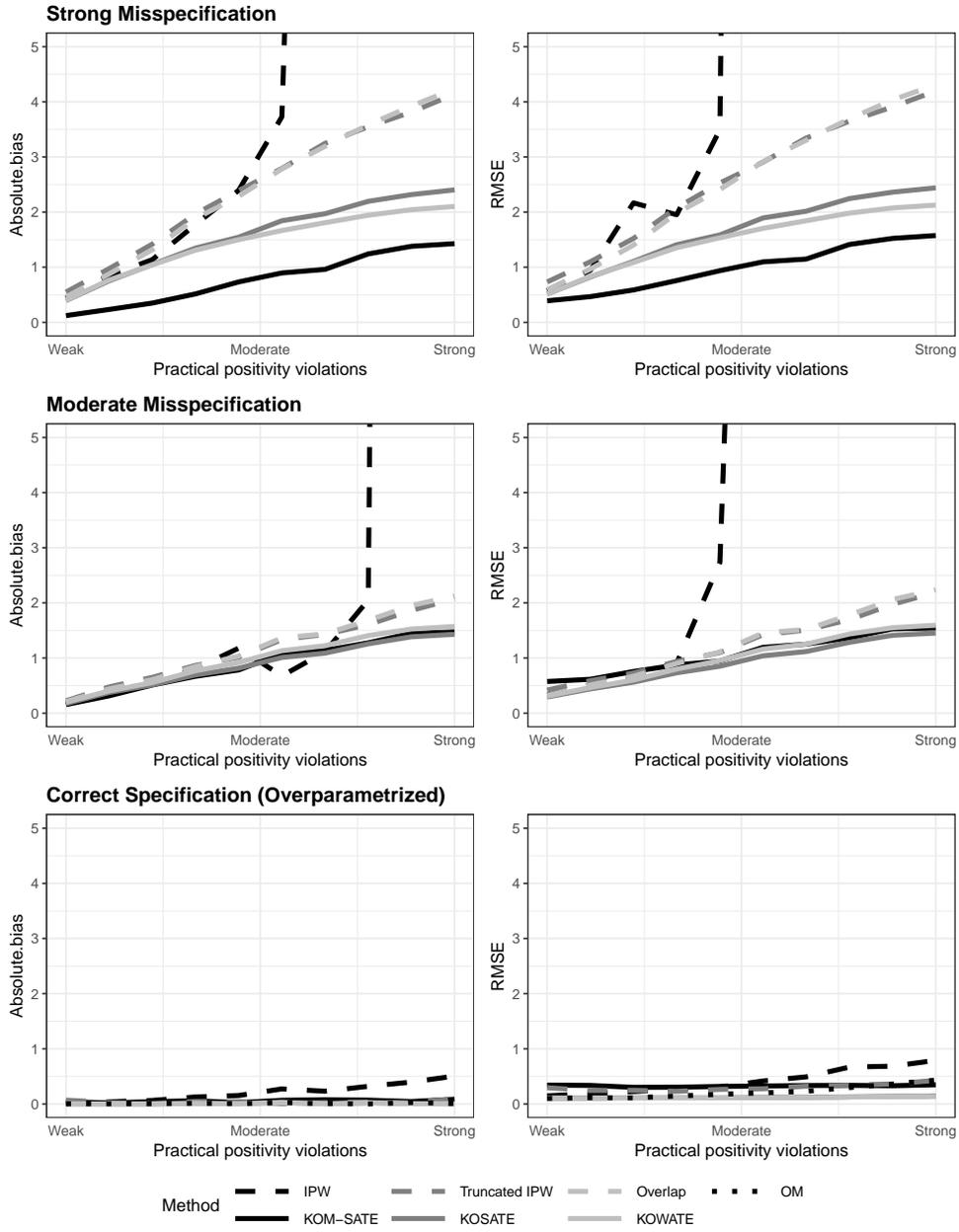}
\end{center}
\caption{\footnotesize ($\lambdas=0$): Absolute bias (left panels) and RMSE (right panels) of SATE estimated by using KOM (solid-black)(which we refer to as KOM-SATE; we will refer to only SATE when clear), KOSATE  by using KOM (solid-dark-grey), KOWATE estimated by using KOM (solid-light-grey), SATE estimated by using IPW (long-dashed-black), OSATE estimated by using truncated weights (long-dashed-dark-grey), OWATE estimated by using overlap weights (long-dashed-light-grey), and SATE estimated by using outcome regression modeling (dashed-light-grey)(which we refer to as OM) when increasing the strength of practical positivity violation under strong misspecification (top panels), moderate misspecification (middle panels) and correct specificiation (overparametrized) (bottom panels), $n=400$.
\label{figowate0} }
\end{figure}

In this Section, we provide additional simulations results. Specifically, we  evaluate the performance of KOM with respect to root MSE 1) across levels of practical positivity violations and across  levels  of  misspecification, when $\lambdas=0$ (Figure \ref{figowate0}), and 2) when increasing sample size and across levels of the penalization parameter $\lambdas=\lambda$ under correct specification (Figure \ref{figowaten}).


\subsection{Estimating GATE when increasing the sample size and the level of $\lambdas$}
In this section we describe the setup for the evaluation of the performance of KOM in estimating GATE when increasing the sample size and the level of the penalization parameter $\lambdas$ under no misspecification. We considered three practical positivity violations: weak, moderate and strong, defined as described in Section \ref{simu}. We considered five different samples sizes, $n=100$ to $n=500$, and thirty different values for $\lambdas=\lambda$, from $0$ to $100$. When evaluating KOM across sample sizes, we set $\lambda_t=\frac{\sigma_t^2}{\gamma_t^2}$, \ie~ the estimated optimal $\lambdas$ for targeting CMSE, while when evaluating KOM across levels of $\lambdas$ we set the sample size equal to $n=100$. For each scenario, we computed the potential outcomes as described in Section \ref{setup} and we estimated SATE, KOSATE and KOWATE by solving optimization problems \eqref{kom_cmse1} and \eqref{kom_cmse2}. We used a polynomial kernel degree 1 (a linear kernel), and plugged into the kernel the correctly specified covariates, $X_1$ and $X_2$.

\subsubsection{Results when increasing sample size and $\lambdas$ under no misspecification}

\begin{figure}[H] 
\begin{center}
\includegraphics[scale=.6]{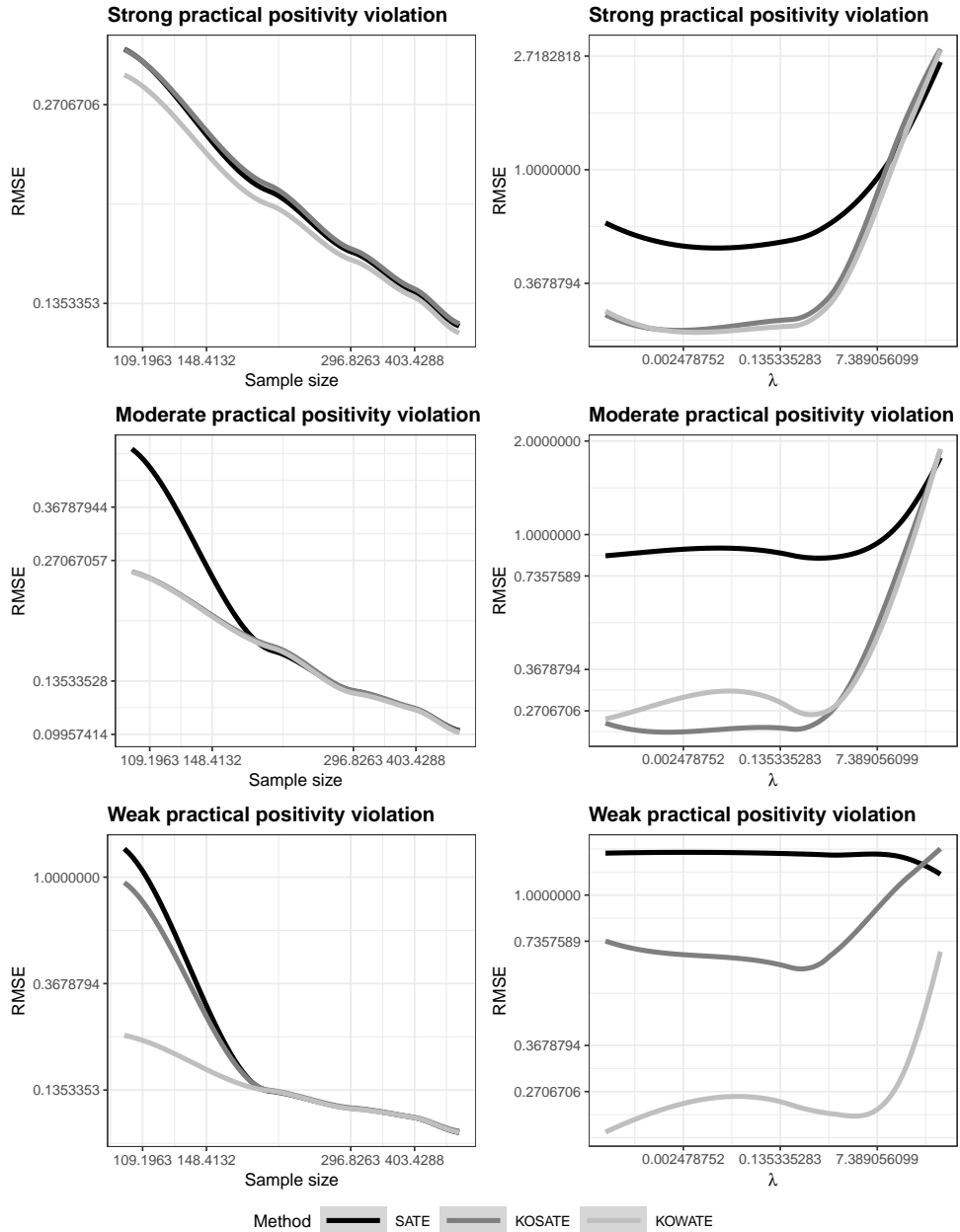}
\end{center}
\caption{\footnotesize Performance of KOM when estimating KOWATE (solid-dark-grey), KOSATE (solid-light-grey) and SATE (solid-black), with respect of RMSE when increasing the sample size with $\lambda_t=\frac{\sigma_t^2}{\gamma_t^2}$ (left panels) and when increasing the penalization parameter $\lambdas=\lambda$ with sample size $n$ set to be equal to 100 (right panels), across weak, moderate and strong practical positivity violation scenarios. Results are presented in the log-transformed scale. 
\label{figowaten} }
\end{figure}

Figure \ref{figowaten} shows the performance of KOM when estimating SATE (solid-black), KOSATE (solid-dark-grey) and KOWATE (solid-light-grey), with respect of RMSE when increasing the sample size with $\lambda_t=\frac{\sigma_t^2}{\gamma_t^2}$ (left panels) and when increasing the penalization parameter $\lambdas=\lambda$ with sample size $n$ set to be equal to 100 (right panels), across strong, moderate and weak practical positivity violation scenarios. Results are presented in the log-transformed scale. 

In summary, RMSEs decreased when increasing the sample size for SATE, KOSATE and KOWATE with a similar rate across weak, moderate and strong practical positivity violation scenarios.  

Lower values of the penalization parameter $\lambdas$ did not seem to affect the performance of KOM with respect of RMSE for KOWATE, KOSATE and SATE across all three practical positivity violation scenarios.

\subsection{Some considerations about standard error estimation and coverage}
\label{se_cove}

 When dealing with weighted estimators, several authors suggested using a robust ``sandwich'' variance estimator \citep{freedman2006so}. Furthermore, to  compute  confidence  intervals  of  a  weighted  estimator,  Wald  confidence intervals can be used \citep{hernan2001marginal,robins2000marginal,freedman2006so}. In Section \ref{gate}, Theorem \ref{thm1}, we showed that the conditional variance of the weighted estimator, $\hat \tau_W$ is equal to the sum of squared weights multiply by the variance of the error, \ie,~$\frac{1}{n^2}\sum_{i=1}^{n}S_i W_i^2\sigma^2$. In this Section we provide some practical considerations about standard error estimation and coverage of the 95\% confidence interval. 

\begin{figure}[H] 
\begin{center}
\includegraphics[scale=.74]{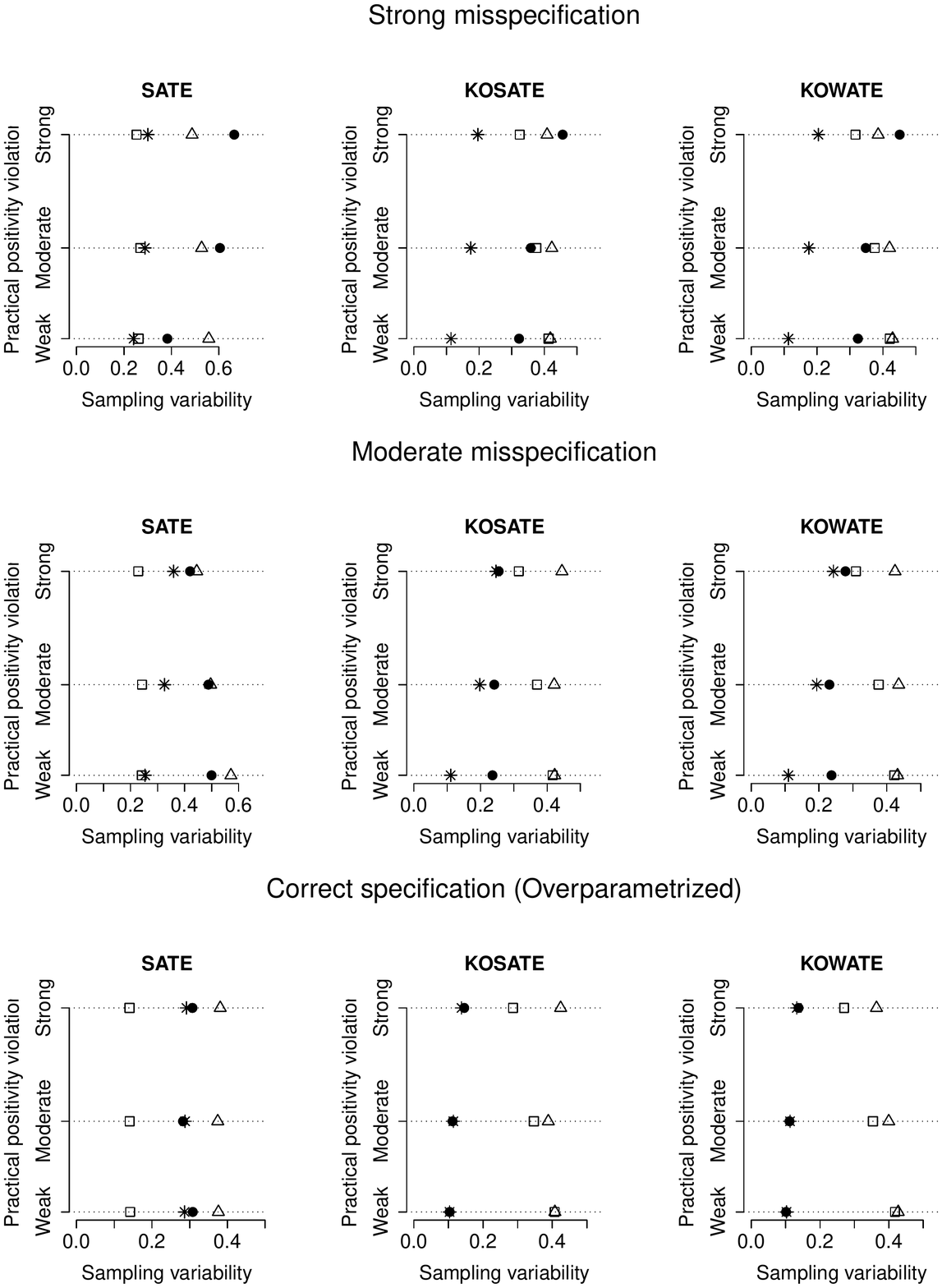}
\end{center}
\caption{\footnotesize \textit{Sampling variability}, $\lambda_t=\frac{\sigma_t^2}{\gamma_t^2}$: Empirical versus estimated standard errors. $\bullet$ empirical standard error of $\hat\tau_W$; $\ast$ conditional standard error; $\square$ naive standard error from OLS; $\triangle$ robust ``sandwich'' standard error. 
\label{figowate_var1} }
\end{figure}

\begin{figure}[H] 
\begin{center}
\includegraphics[scale=.74]{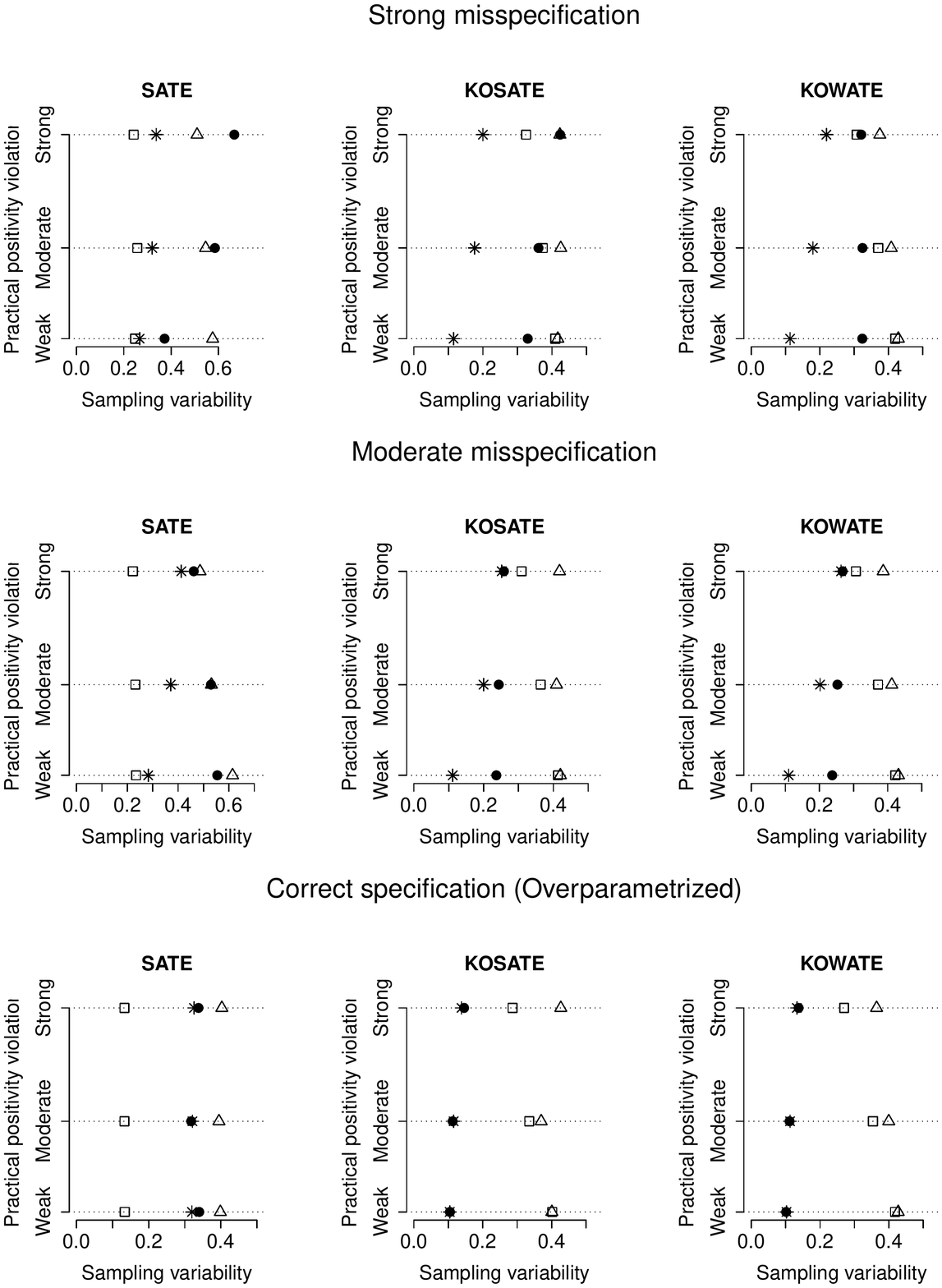}
\end{center}
\caption{\footnotesize \textit{Sampling variability}, $\lambdas=0$: Empirical versus estimated standard errors. $\bullet$ empirical standard error of $\hat\tau_W$; $\ast$ conditional standard error; $\square$ naive standard error from OLS; $\triangle$ robust ``sandwich'' standard errror. 
\label{figowate_var0} }
\end{figure}

\begin{figure}[H]
\begin{center}
\includegraphics[scale=.74]{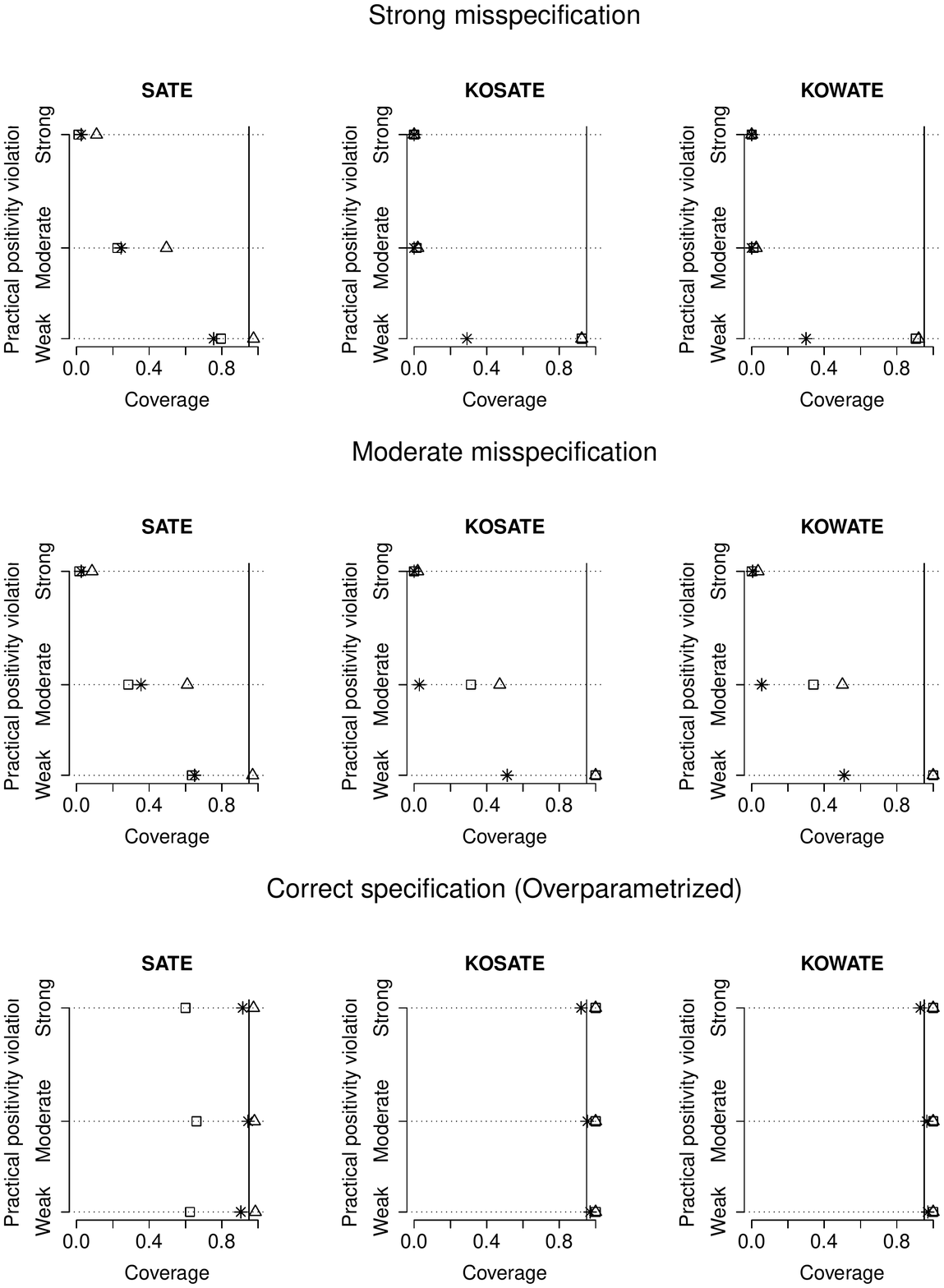}
\end{center}
\caption{\footnotesize \textit{Coverage}, $\lambda_t=\frac{\sigma_t^2}{\gamma_t^2}$: Coverage if the 95\% confidence interval (CI). $\ast$ coverage by using conditional standard error; $\square$ coverage by using naive standard error from OLS; $\triangle$ coverage by using robust ``sandwich'' standard errror. 
\label{figowate_cove1} }
\end{figure}

\begin{figure}[H] 
\begin{center}
\includegraphics[scale=.74]{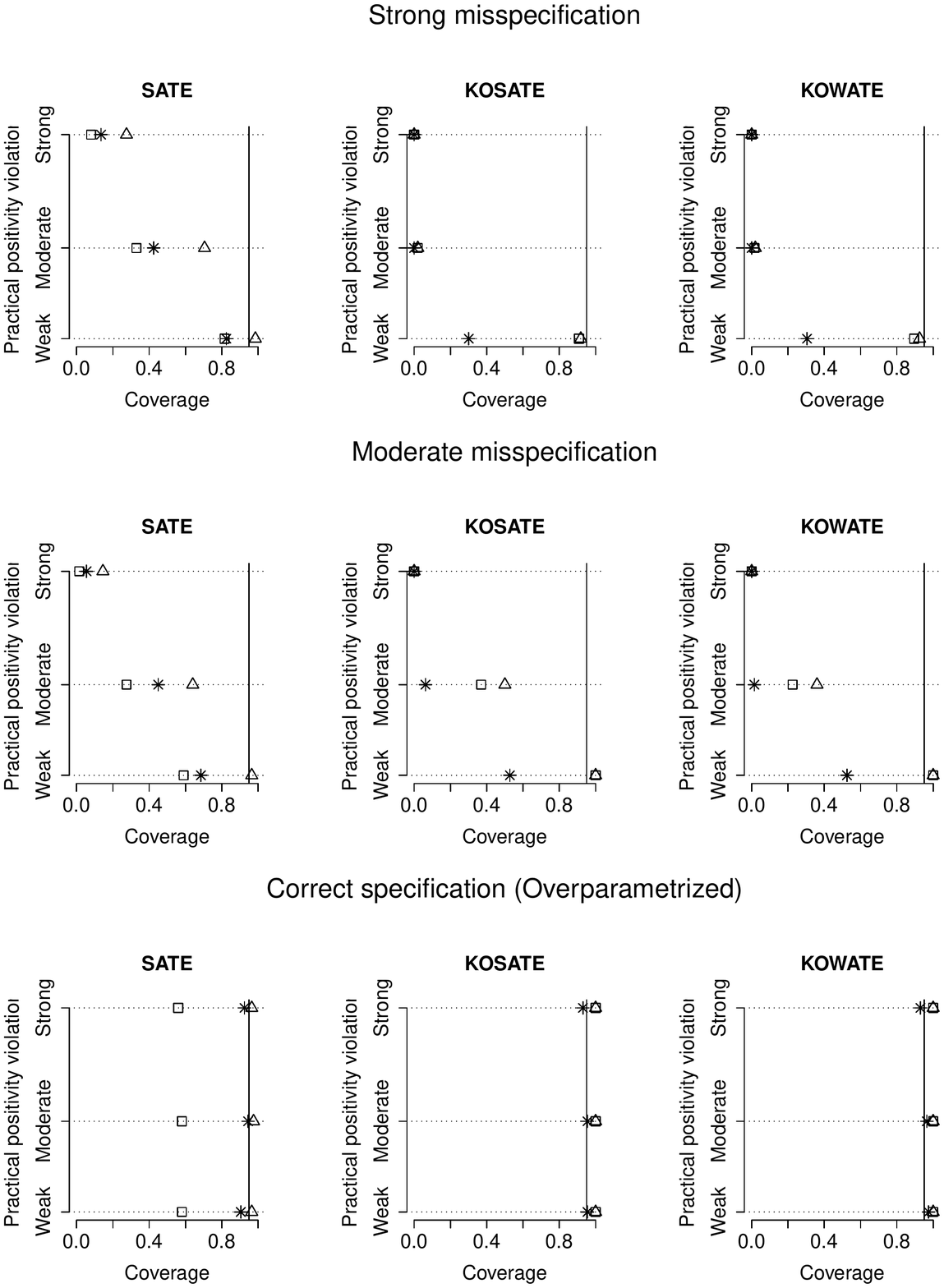}
\end{center}
\caption{\footnotesize \textit{Coverage}, $\lambdas=0$: Coverage if the 95\% confidence interval (CI). $\ast$ coverage by using conditional standard error; $\square$ coverage by using naive standard error from OLS; $\triangle$ coverage by using robust ``sandwich'' standard errror. 
\label{figowate_cove0} }
\end{figure}

Figures \ref{figowate_var1} and  \ref{figowate_var0} show the results of a simulation study aimed at comparing the empirical standard error of the sampling distribution of estimated SATE, KOSATE and KOWATE when $\lambda_t=\frac{\sigma_t^2}{\gamma_t^2}$ and $\lambdas=0$ respectively, with 1) the standard error obtained in Theorem \ref{thm1} (conditional standard error), 2) the classic standard error obtained by using ordinary least squares under errors homoskedasticity, and 3) the robust ``sandwich'' standard error, across levels of practical positivity violations under strong misspecification (top panels), moderate misspecification (middle panels) and correct specification (bottom panels). We used the \textsf{R} package \textsf{sandwich} to estimate robust standard errors. Similarly, Figures \ref{figowate_cove1} and \ref{figowate_cove0} the coverage of the 95\% Wald confidence interval across levels of practical positivity violations when estimating SATE, KOSATE and KOWATE.  

In summary, under strong and moderate misspecification, the conditional standard error was smaller than the empirical standard error for SATE, KOSATE and KOWATE. The sandwich estimator was larger, across almost all scenarios of practical positivity violations for SATE, KOSATE and KOWATE.  The naive standard error was slightly larger than the empirical standard error for KOSATE and KOWATE, while smaller for SATE. Under correct specification,  the conditional standard error was very close to the empirical standard error for all the methods.  We obtained similar results for $\lambdas=0$. 

Under strong and moderate misspecification the coverage of the 95\% confidence interval was close to nominal values when using the ``sandwich'' estimator under weak practical positivity violations. For moderate and strong practical positivity violations the coverage was low for all methods and for both the estimated optimal $\lambdas$ and $\lambdas=0$. As shown in \cite{kallus2016generalized}, exact coverage can be obtained by using the conditional standard error when computed keeping $\Xs,\Ts$ fixed under no model misspecification. Finally, in practical settings, we suggest to use Wald confidence intervals together with the robust ``sandwich'' standard error estimator as previously suggested by other authors \citep{hernan2001marginal,robins2000marginal,freedman2006so}.

\subsection{Computational time of KOM}

To find a solution to the optimization problems \eqref{qpkom1}, and \eqref{qpkom2}, three steps are required: (1) tune the kernel hyperparameters; (2) compute the matrices and (3) solve the quadratic optimization problem. In this Section we provide a summary about the computational cost of finding a solution. We computed the computational time by using the \textsf{R} package \textsf{rbenchmark} on a AWS EC2 C5 instance, Intel Xeon Platinum 8000 series, 3.5 GHz, 32GB RAM and a Linux Ubuntu 16.04 operating system. Table \ref{tab_ct_owate} shows the mean computational time in seconds needed to tune the hyperparameters across treated (GPML 1) and control (GPML 0), constructing the matrices (Matrices) and solving the optimization problem with \textsf{Gurobi} (Gurobi) when estimating KOWATE, SATE, and KOSATE across the positivity and misspecified scenarios previously described. In summary, for KOSATE and KOWATE most of the computational time was needed to solve the optimization problem, especially for KOSATE. 

On average, SATE and KOWATE with a product of polynomial degree 2 always found a solution across levels of practical positivity violations and across levels of misspecification. KOWATE with a product of polynomial degree 2 found a solution around 71\% of the times, while the remaining by using a polynomial degree 3. We argue that this is due to the different formulation of the optimization problem, which includes binary variable in the constraint.


\begin{table}[H]
\centering
\caption{Average total time for each of the processes needed to obtain the set of KOM weights. \label{tab_ct_owate}}
\begin{tabular}{ccccccc}
\hline
              & \multicolumn{3}{c}{\textbf{Practical Positivity Violation}} \\ \cline{2-4} 
              &  SATE &  KOSATE & KOWATE  \\ \cline{2-4} 
\textbf{Task} &  Seconds      &         &     \\ \cline{1-1}
GPML 0        &  0.58   & 0.50     & 0.58           \\
GPML 1        & 0.28  & 0.28    & 0.28          \\
Matrices    & 0.89   & 0.90     & 0.92          \\ 
Gurobi        & 0.2  & 22.14    & 6.46               \\
\hline
\end{tabular}
\end{table}

\newpage

\section{Omitted proofs}

\subsection{Proof of Lemma \ref{lemma1}}

Note that, under assumption \ref{honest}, to obtain a set of weights $\Ws$ that makes $\hat \tau_{W}$ unbiased for $\hat \tau_V$ what we need is $\Eb{S_i \indic{T_i=t}W_i | \Xs, T_{-i}, \Ss}$ to be equal to $\Eb{V_i | \Xs, T_{-i}, \Ss}$ for each $t \in {0,1}$, where $T_{-i}$ is equal to $\Ts$ except for the $i$-th unit (we refer to $(T_{-i},t)$ as equal to $\Ts$ except for the $i$-th unit which is set to be equal to $t$). Note that
\begin{align*}
\Eb{S_i \indic{T_i=1}W_i | \Xs, T_{-i}, \Ss} &= \phi(X_i) \psi(X_i)W_i(\Xs,(T_{-i},1),\Ss) \\
\Eb{S_i \indic{T_i=0}W_i | \Xs, T_{-i}, \Ss} &= (1-\phi(X_i)) \psi(X_i)W_i(\Xs,(T_{-i},0),\Ss) \\
\Eb{V_i | \Xs, T_{-i}, \Ss}&=\phi(X_i)V_i(\Xs,(T_{-i},1),\Ss) + (1-\phi(X_i))V_i(\Xs,(T_{-i},0),\Ss) \\
\end{align*}

Then, what we want is 
\begin{align*}
    \phi(X_i) \psi(X_i)W_i(\Xs,(T_{-i},1),\Ss)&=\phi(X_i)V_i(\Xs,(T_{-i},1),\Ss) \\&+ (1-\phi(X_i))V_i(\Xs,(T_{-i},0),\Ss) \\
    (1-\phi(X_i)) \psi(X_i)W_i(\Xs,(T_{-i},0),\Ss)&=\phi(X_i)V_i(\Xs,(T_{-i},1),\Ss) \\&+ (1-\phi(X_i))V_i(\Xs,(T_{-i},0),\Ss).
\end{align*}

Solving for $W_i(\Xs,(T_{-i},t),\Ss)$ for each $t$ leads to the weights 
\begin{align*}
    W_i(\Xs,(T_{-i},1),\Ss)&= \prns{\frac{T_i}{\phi(X_i) \psi(X_i)}} ( \phi(X_i)V_i(\Xs,(T_{-i},1),\Ss) \\&+ (1-\phi(X_i))V_i(\Xs,(T_{-i},0),\Ss) )\\
    W_i(\Xs,(T_{-i},0),\Ss)&= \prns{\frac{(1-T_i)}{(1-\phi(X_i)) \psi(X_i)}} (\phi(X_i)V_i(\Xs,(T_{-i},1),\Ss) \\&+ (1-\phi(X_i))V_i(\Xs,(T_{-i},0),\Ss)).
\end{align*}

Combining the two set of weights we obtain the weights $W_i^{\op{IPW}}$ as shown in Lemma \ref{lemma1}.

\subsection{Proof of Theorem \ref{thm1}}

Bias of $\hat{\tau}_W$ with respect to $\tau_V$. We show that 
 
\begin{equation}
\begin{aligned}
\Eb{\hat{\tau}_W - \tau_V |  H_{1:n} } &=  B_1(\Ws,\Vs,g_1) - B_0(\Ws,\Vs,g_0)
\end{aligned}
\end{equation}

Define $\epsilon_{i,t} = Y_i(t)-g_t(X_i)$, $\epsilon_{i}=T_i\epsilon_{i,1}+(1-T_i)\epsilon_{i,0}$, and
 
\begin{equation}
  \Xi(\Ws) = \frac{1}{n} \sum_{i = 1}^n S_i  T_i W_i  \epsilon_{i,1}  - \frac{1}{n} \sum_{i = 1}^n S_i  (1-T_i) W_i  \epsilon_{i,0}.
\end{equation}

Then, 
\begin{equation}
\label{biasp}
\begin{aligned}
	\hat{\tau}_W - \tau_V  &=  \frac{1}{n} \sum_{i = 1}^n S_i W_i ( T_i Y_i - (1-T_i) Y_i ) - \frac{1}{n}\sum_{i = 1}^n V_i (g_1(X_i)-g_0(X_i)) \\
		&= \frac{1}{n} \sum_{i = 1}^n  (S_i T_i W_i  - V_i ) g_1(X_i) - \frac{1}{n} \sum_{i = 1}^n  (S_i (1-T_i) W_i  - V_i ) g_0(X_i) \\ 
		&\phantom{=}+ \frac{1}{n} \sum_{i = 1}^n S_i   T_i W_i \epsilon_{i,1}  - \frac{1}{n} \sum_{i = 1}^n S_i (1-T_i) W_i \epsilon_{i,0} \\
	&= B_1(\Ws,\Vs,g_1) - B_0(\Ws,\Vs,g_0) + \Xi(\Ws),
\end{aligned}
\end{equation}

\noindent
where the second equality follows by consistency and by definition of $g_t(X_i)$. For each $i$ and each $t$, by the definition of $g_t$ and by ignorability, $\Eb{ \epsilon_{i,t} | \Xs,\Ts,\Ss} = \Eb{ Y_i(t) | X_{i},T_i,S_i}-g_t(X_i) = \Eb{ Y_i(t) | X_{i}]-g_t(X_i)} = 0.$  By assuming $\Ws$ to be a function of $H_{1:n}$, we have that $\Eb{\Xi(W)\mid H_{1:n}} = 0$ and hence
\begin{equation}
\Eb{\hat{\tau}_W - \tau_V |  H_{1:n} } =  B_1(\Ws,\Vs,g_1) - B_0(\Ws,\Vs,g_0).
\end{equation}


We now show that
\begin{equation}
\begin{aligned}
\Eb{\left( {\hat{\tau}_W - \tau_V} \right)^2 |  H_{1:n} } = \left(B_1(\Ws,\Vs,g_1) - B_0(\Ws,\Vs,g_0)\right)^2 + \frac{1}{n^2}\sum_{i = 1}^n S_i^2W_i^2 \sigma_i^2. 
\end{aligned}
\end{equation}

We define $\sigma^2_{i,t}=\text{Var}({Y_i(t)\mid X_i, S_i})$. Under consistency, non-interference and ignorability, we have $\sigma^2_{i}=\text{Var}({Y_i\mid H_i})=S_iT_i\sigma_{i,1}^2+S_i(1-T_i)\sigma_{i,0}^2$.
We can decompose the CMSE as bias squared plus variance.  To compute the variance we consider only $\Xi$. Then, for each $i,j$, we have $\Eb {S_iS_j  W_iW_j (-1)^{T_i+T_j}  \epsilon_{i} \epsilon_{j} |  H_{1:n} } = S_iS_j W_iW_j (-1)^{T_i+T_j}\Eb { \epsilon_{i} \epsilon_{j} |  H_{1:n}}$. When $i \neq j$, $S_iS_jW_iW_j (-1)^{T_i+T_j}\Eb { \epsilon_{i} \epsilon_{j} |  H_{1:n} }= S_iS_jW_iW_j (-1)^{T_i+T_j} \Eb { \epsilon_{i} |  H_{1:n}  }\Eb { \epsilon_{j} |  H_{1:n}  } = 0$. 
When $i = j$,  $S_iS_jW_iW_j (-1)^{T_i+T_j} \Eb { \epsilon_{i} \epsilon_{j} |  H_{1:n}  } = S_i^2W_i^2 \sigma_i^2$.

\subsection{Proof of Theorem \ref{thm2}}

Define the matrix $K_t\in\mathbb R^{n\times n}$ as $K_{tij}=\mathcal K_t(X_{i},X_{j})$ (that such a matrix is PSD for any set of points is precisely the definition of a PSD kernel). By the representer theorem, we have that
\begin{align*}
\Delta^2_t(\Ws,\Vs) &=  \sup_{\|g\|^2_{t} \leq 1} \prns{ \frac{1}{n} \sum_{i = 1}^n  \prns{\mathbbm{1}[S_i=s] \mathbbm{1}[T_i=t]W_i - V_i } g_t(X_i)}^2 \\
&= \sup_{\sum_{i,j=1}^n\alpha_i\alpha_j\mathcal K_t(X_i,X_j)\leq1} \prns{ \frac{1}{n} \sum_{i = 1}^n  \prns{\mathbbm{1}[S_i=s] \mathbbm{1}[T_i=t]W_i - V_i } \sum_{j=1}^n\alpha_j\mathcal K_t(X_i,X_j)}^2\\
&= \sup_{\alpha^TK_t\alpha\leq1} \prns{\frac{1}{n}\alpha^TK_t(I_SI_t\Ws-\Vs)}^2\\
&=\frac{1}{n^2}(I_SI_{t}\Ws-\Vs)^TK_t(I_SI_{t}\Ws-\Vs) \\
&=\frac{1}{n^2} \prns{\Ws^TI_SI_{t}K_tI_SI_{t}\Ws-2\Vs^TK_tI_SI_{t}\Ws+\Vs^TK_t\Vs}.
\end{align*}

\newpage
\subsection{Proof of Theorem \ref{thm3}}

We start proving Theorem \ref{thm3} by showing that the worst-case CMSE of the weighted estimator, weighted by the normalized IPW weights is $O_p(1/n)$. Recall that the worst-case CMSE defined as
\begin{align}
\label{worstCMSE}
\mathfrak{C}(\Ws,\Vs,\lambdas) &=\Delta^2_1(\Ws,\Vs)+\Delta^2_0(\Ws,\Vs)+\frac{\lambda_0}{n^2} \| I_sI_0 \Ws \|_2^2 +\frac{\lambda_1}{n^2} \| I_sI_1 \Ws \|_2^2 ,
\end{align}

Based on the development of $W_i^{\op{IPW}}$ showed in Lemma \ref{lemma1} consider for each $t \in \lbrace 0,1 \rbrace$ the normalized IPW weights $W_i^{\op{nIPW}}=\frac{W_i^{\op{IPW}}}{Z_{T_i}^{\op{IPW}}}$ where $Z_t^{\op{IPW}}=\frac1n\sum_{i \in \mathcal{T}_t} W_i^{\op{IPW}}$, $\mathcal{T}_1 = \TT$ and $\mathcal{T}_0 = \TC$. Consider also a generic dummy function $f$. Then, for each $t$ we have
\begin{align*}
    \Delta_t(W_{1:n}^{\op{nIPW}},\Vs) &= \sup_{\|f\|_{t} \leq 1} B_t(W_{1:n}^{\op{nIPW}},\Vs,f) \\
    &= \frac{1}{Z_t^{\op{IPW}}} \sup_{ \| f \|_{t} \leq 1} \frac{1}{n} \sum_{i=1}^n f(X_i) \prns{S_i\indic{T_i=t}W_i^{\op{IPW}} - V_iZ_t^{\op{IPW}} } \\
    &\leq \underbrace{\frac{1}{Z_t^{\op{IPW}}} \sup_{ \| f \|_{t} \leq 1} \frac{1}{n} \sum_{i=1}^n f(X_i) \prns{S_i\indic{T_i=t}W_i^{\op{IPW}} - V_i }}_{(a)} \\
    &+  \underbrace{\frac{1}{Z_t^{\op{IPW}}} \sup_{ \| f \|_{t} \leq 1} \frac{1}{n} \sum_{i=1}^n f(X_i) \prns{Z_t^{\op{IPW}}-1 }V_i }_{(b)} &&\text{(Triangle inequality)}
\end{align*}

We now show that $(a)$ is $O_p(1/n)$. Let $\xi_i(f)=f(X_i) \prns{S_i\indic{T_i=t}W_i^{\op{IPW}} - V_i }$, and note that $\Eb{ \Eb{\xi_i | \Xs, \Ss}}=0$ for all $f$ and $i$. We can now therefore use the symmetrization trick. Then, let $\xi_i'$ for all $i=1, \dots n$, be iid replicates of $\xi_i$ and let $\rho_i$ be iid Rademacher random variables independent of all else, we have
\begin{align*}
    \Eb{ \prns{ \sup_{ \| f \|_{t} \leq 1} \frac{1}{n} \sum_{i=1}^n \xi_i(f) }^2} &= \Eb{\| \frac{1}{n} \sum_{i=1}^n \xi_i(f) \|_{\mathcal{K}_t}^2} \\
    &=\frac{1}{n^2}\Eb{\| \sum_{i=1}^n \Eb{\xi_i'(f)} - \xi_i(f) \|_{\mathcal{K}_t}^2} &&(\Eb{\xi_i'(f)}=0) \\
    &\leq \frac{1}{n^2}\Eb{\| \sum_{i=1}^n \xi_i'(f) - \xi_i(f) \|_{\mathcal{K}_t}^2} &&(\text{Jensen's inequality}) \\
    &= \frac{1}{n^2}\Eb{\| \sum_{i=1}^n \rho_i \prns{\xi_i'(f) - \xi_i(f)} \|_{\mathcal{K}_t}^2} &&(\text{Rademacher rvs}) \\
    &\leq \frac{4}{n^2}\Eb{\| \sum_{i=1}^n \rho_i \xi_i(f) \|_{\mathcal{K}_t}^2}. &&(\text{Triangle inequality})
\end{align*}

Note that $\| \xi_1(f) - \xi_2(f) \|^2_{\mathcal{K}_t} + \| \xi_1(f) + \xi_2(f) \|^2_{\mathcal{K}_t}=2\| \xi_1(f)  \|^2_{\mathcal{K}_t}+2\|  \xi_2(f) \|^2_{\mathcal{K}_t} + 2\langle \xi_1(f),\xi_2(f) \rangle - 2\langle \xi_1(f),\xi_2(f) \rangle = 2\| \xi_1(f)  \|^2_{\mathcal{K}_t}+2\|  \xi_2(f) \|^2_{\mathcal{K}_t}$. Then by induction, $\sum_{\rho_i \in \lbrace -1,1 \rbrace}^n \| \rho_i \xi_i(f) \|^2_{\mathcal{K}_t}=2^n\sum_{i=1}^n\| \xi_i(f) \|^2_{\mathcal{K}_t}$. 

By Assumptions~\ref{bound1} and \ref{bound2}, $0 < \eta_{\phi} < \phi < 1-\eta_{\phi}$ and $0 < \eta_{\psi} < \psi < 1-\eta_{\psi}$  for some $\eta_\phi,\eta_\psi$.
Therefore, 
\begin{align*}
    \| \xi_i(f) \|^2_{\mathcal{K}_t} &\leq \prns{S_i\indic{T_i=t}W_i^{\op{IPW}} - V_i }^2\mathcal{K}_t(X_i,X_i) 
    \leq \prns{\frac{2}{\eta_{\psi}^2 \eta_{\phi}^2} + 2V_i^2}\mathcal{K}_t(X_i,X_i), 
\end{align*}
\noindent
where the first inequality follows by the reproducing property and from the fact that $\prns{S_i\indic{T_i=t}W_i^{\op{IPW}} - V_i }$ is a constant, while the second inequality follows from the fact that $W_i^{\op{IPW}}$ squared is bounded by $1/\eta_{\psi}^2 \eta_{\phi}^2$ by assumptions \ref{bound1} and \ref{bound2}. Consequently, 
\begin{align*}
    \Eb{ \| \xi_i(f) \|^2_{\mathcal{K}_t}} &\leq  \underbrace{\frac{2}{\eta_{\psi}^2 \eta_{\phi}^2} \Eb{\mathcal{K}_t(X_i,X_i)}}_{(a1)} + \underbrace{ 2\Eb{V_i^2\mathcal{K}_t(X_i,X_i)}}_{(a2)}. 
\end{align*}

By assumptions \ref{bound1} and \ref{bound2}, and since we are using bounded kernels, $(a1) < \infty$. Furthermore, by assuming $\Eb{V_i^2} < \infty$ we have that $(a2) < \infty$ and consequently we get $\Eb{\| \frac{1}{n} \sum_{i=1}^n \xi_i(f) \|_{\mathcal{K}_t}^2} = O(1/n)$ and by Markov's inequality $\Eb{\| \frac{1}{n} \sum_{i=1}^n \xi_i(f) \|_{\mathcal{K}_t}^2} = O_p(1/n)$. Then, $(a)=O_p(1/n)$. We now evaluate $(b)$. Note that 
\begin{align*}
    (b) &= \frac{1}{Z_t^{\op{IPW}}} \sup_{ \| f \|_{t} \leq 1} \frac{1}{n} \sum_{i=1}^n f(X_i) \prns{Z_t^{\op{IPW}}-1 } V_i \\
    &= \frac{1}{Z_t^{\op{IPW}}} \sup_{ \| f \|_{t} \leq 1} \frac{1}{n} \sum_{i=1}^n \prns{ \langle \mathcal{K}_t(X_i,\cdot),f \rangle } \prns{Z_t^{\op{IPW}}-1 } V_i &&\text{(reproducing property)} \\
    &= \frac{1}{Z_t^{\op{IPW}}} \| \frac{1}{n} \sum_{i=1}^n \prns{ \langle \mathcal{K}(X_i,\cdot),f \rangle } \prns{Z_t^{\op{IPW}}-1 } V_i \|_{\mathcal{K}_t} &&\text{(definition of dual norm)} \\
    &= \frac{1}{Z_t^{\op{IPW}}} \mid \prns{Z_t^{\op{IPW}}-1 } \mid \| \frac{1}{n} \sum_{i=1}^n \prns{ \langle \mathcal{K}(X_i,\cdot),f \rangle }  V_i \|_{\mathcal{K}_t} &&\text{(absolutely homogeneous property)} \\
    &\leq \frac{1}{Z_t^{\op{IPW}}} \abs{\prns{Z_t^{\op{IPW}}-1 } }  \frac{1}{n} \sqrt{ \sum_{i=1}^n \langle \mathcal{K}(X_i,\cdot),f \rangle^2} \sqrt{ \sum_{i=1}^n V_i^2} &&\text{(Cauchy-Schwarz inequality)} \\
    &= \underbrace{\frac{1}{Z_t^{\op{IPW}}}  \abs{\prns{Z_t^{\op{IPW}}-1} }}_{(b1)} \underbrace{ \frac{1}{n} \sum_{i=1}^n \sqrt{    \mathcal{K}_t(X_i,X_i)}}_{(b2)} \underbrace{ \sqrt{ \sum_{i=1}^n V_i^2}.}_{(b3)} &&\text{(reproducing property)}
\end{align*}

We now show that $(b)$ is $O_p(1/n)$. We evaluate $(b1)$ first. 
Note that by assumptions \ref{bound1} and \ref{bound2} and since $\Eb{ W_{i}^{\op{nIPW}} }=1$ and $\Eb{ (W_{i}^{\op{nIPW}})^2 }<\infty$, then $\Eb{(Z_t^{\op{IPW}}-1)^2}=O(1/n)$ and by Markov's inequality, $(Z_t^{\op{IPW}}-1)^2=O_p(1/n)$. We now evaluate $(b2)$. Since we assumed bounded kernels, we have that $\Eb{\sqrt{\mathcal{K}_t(X_i,X_i)}}<\infty$, and by the law of large number we have that $\frac{1}{n}\sum_{i=1}^n  \sqrt{\mathcal{K}_t(X_i,X_i)} \xrightarrow{p} \Eb{\sqrt{\mathcal{K}_t(X_i,X_i)} }$ and therefore $(b2)=O_p(1)$. We now evaluate $(b3)$. Recall that $\Vs \in \mathcal V=\fbraces{\Vs\in\R n:V_i\geq0\;\forall i,\sum_{i}^nV_i=1}$. Note that  $\sqrt{ \sum_{i=1}^n V_i^2}\leq \sum_{i=1}^n V_i$ and $\sum_{i=1}^n V_i = 1$ by definition of $\mathcal V$. Then, by assuming $\Eb{V^2}<\infty$, then  $\sqrt{ \sum_{i=1}^n V_i^2}\leq \sum_{i=1}^n V_i=O(1)$ and by Markov's inequality is $O_p(1)$. Finally, combining all together, by the Slutsky's theorem we have that $(b)=O_p(1/n)$. Consequently, we have that $\Delta^2_1(\Ws,\Vs)+\Delta^2_0(\Ws,\Vs)=O_p(1/n)$.

We now focus on the third term of the worst case CMSE, the conditional variance of the weighted estimator. By assuming $\sigma_i^2 < \infty$ and and assumptions \ref{bound1} and \ref{bound2}, $\| \frac{1}{n^2} W_{1:n}^{\op{nIPW}} \|_2^2=O(1/n)$, and by Markov's inequality $\frac{1}{n^2}\| W_{1:n}^{\op{nIPW}} \|_2^2=O_p(1/n)$. By the Slutsky's theorem combining it with $(a)$ and $(b)$ we have that, $\mathbbm{E}\left[ \prns{\hat{\tau}_{W_{1:n}^{\op{nIPW}}} - \tau_V}^2 \given  H_{1:n} \right]=O_p(1/n)$.

Define $W_{1:n}^*$ the set of weights solution to the optimization problem \eqref{kom_cmse1} and recall that $W_{1:n}^{\op{nIPW}} \geq 0$, $e_n^TI_{S}I_1W_{1:n}^{\op{nIPW}}=e_n^TI_{S}I_0W_{1:n}^{\op{nIPW}}=n$ . Then, 
\begin{align*}
    \mathfrak{C}(W_{1:n}^*,\Vs,\lambdas) &\leq \mathfrak{C}(W_{1:n}^{\op{nIPW}},\Vs,\lambdas)  \\
    &\leq\mathfrak{K}_1^2 \Delta^2_1(\Ws,\Vs)+\mathfrak{K}_0^2\Delta^2_0(\Ws,\Vs)+\frac{\mathfrak{K}}{n^2} \| I_s \Ws \|_2^2=O_p(1/n).
\end{align*}
Therefore, 
\begin{align*}
    \Delta^2_t(W_{1:n}^*,\Vs) &\leq \mathfrak{K}_t^{-2} \mathfrak{C}(W_{1:n}^*,\Vs,\lambdas) = O_p(1/n) \\
    \frac{1}{n^2} \| I_s \Ws \|_2^2&\leq\mathfrak{K}^{-1} \mathfrak{C}(W_{1:n}^*,\Vs,\lambdas)=O_p(1/n).
\end{align*}
Then, we have that $\mathbbm{E}\left[ \prns{\hat{\tau}_{W^*_{1:n}} - \tau_V}^2 \given  H_{1:n} \right] = O_p(1/n)$. Finally, from Lemma 31 of \cite{kallus2016generalized}, we have that $\hat{\tau}_{W^*_{1:n}} - \tau_V = O_p(1/\sqrt n)$.

\end{document}